\documentclass[conference]{IEEEtran}
\IEEEoverridecommandlockouts
\usepackage{cite}
\usepackage{amsmath,amssymb,amsfonts}
\usepackage{algorithmic}
\usepackage{graphicx}
\usepackage{textcomp}
\usepackage{xcolor}
\usepackage{hyperref} 
\usepackage{tabularx} 
\usepackage{footnote} 
\usepackage{svg} 
\usepackage{subcaption} 
\usepackage[textsize=tiny]{todonotes} 
\usepackage{cleveref} 
\makesavenoteenv{tabularx} 
\renewcommand\IEEEkeywordsname{Keywords} 

\newcolumntype{C}{>{\centering\arraybackslash}X} 

\def\BibTeX{{\rm B\kern-.05em{\sc i\kern-.025em b}\kern-.08em
    T\kern-.1667em\lower.7ex\hbox{E}\kern-.125emX}}
\begin{document}

\title{How Much Data is Enough? Optimization of Data Collection for Artifact Detection in EEG Recordings\\ 
}

\author{
   \IEEEauthorblockN{
       Lu Wang-Nöth\IEEEauthorrefmark{1}\IEEEauthorrefmark{2}, Philipp Heiler\IEEEauthorrefmark{1}, Hai Huang\IEEEauthorrefmark{2}, Daniel Lichtenstern\IEEEauthorrefmark{1}, Alexandra Reichenbach\IEEEauthorrefmark{3}, Luis Flacke\IEEEauthorrefmark{1}$^{,1}$ \thanks{$^{1}$Currently: Independent researcher.},\\  Linus Maisch\IEEEauthorrefmark{1}, Helmut Mayer\IEEEauthorrefmark{2}
   }
    \IEEEauthorblockA{\IEEEauthorrefmark{1} brainboost GmbH, Augsburgerstraße 4, 80337 Munich, Germany \\ Email: lu.wang-noeth@protonmail.com}
    \IEEEauthorblockA{\IEEEauthorrefmark{2} Institute for Applied Computer Science, \\ Bundeswehr University Munich, Werner-Heisenberg-Weg 39, 85579 Neubiberg, Germany}
    \IEEEauthorblockA{\IEEEauthorrefmark{3} Center for Machine Learning, Heilbronn University, Max-Planck-Str. 39, 74081 Heilbronn, Germany}
}

\maketitle

\begin{abstract}

    \textit{Objective.} Electroencephalography (EEG) is a widely used neuroimaging technique known for its cost-effectiveness and user-friendliness. However, the presence of various artifacts, particularly biological artifacts like Electromyography (EMG) signals, leads to a poor signal-to-noise ratio, limiting the precision of analyses and applications. The currently reported EEG data cleaning performance largely depends on the data used for validation, and in the case of machine learning approaches, also on the data used for training. The data are typically gathered either by recruiting subjects to perform specific artifact tasks or by integrating existing datasets. Prevailing approaches, however, tend to rely on intuitive, concept-oriented data collection with minimal justification for the selection of artifacts and their quantities. 
    Given the substantial costs associated with biological data collection and the pressing need for effective data utilization, we propose an optimization procedure for data-oriented data collection design using deep learning-based artifact detection. 
    \textit{Approach.} We apply a binary classification between artifact epochs (time intervals containing artifacts) and non-artifact epochs (time intervals containing no artifact) using three different neural architectures. Our aim is to minimize data collection efforts while preserving the cleaning efficiency. 
    \textit{Main results.} We were able to reduce the number of artifact tasks from twelve to three and decrease repetitions of isometric contraction tasks from ten to three or sometimes even just one. \textit{Significance.} Our work addresses the need for effective data utilization in biological data collection, offering a systematic and dynamic quantitative approach. By providing clear justifications for the choices of artifacts and their quantity, we aim to guide future studies toward more effective and economical data collection in EEG and EMG research.
    
\end{abstract}


\section{Introduction}
    Electroencephalography (EEG) is a widely used method of collecting brain activity data for medical and neuroscience research as well as for clinical application. It offers a unique set of advantages including its non-invasive nature, compatibility with subjects in motion, and superior temporal resolution. Furthermore, its cost-effectiveness and user-friendly features render it accessible to both hospital settings and practitioners in general as well as psychotherapists.   
    
    The biggest challenge in working with EEG data lies in addressing the poor signal-to-noise ratio. Each EEG electrode collects the sum of all electrical signals projected onto the scalp, some of which not originating from cortical activity, thus, introducing artifacts into the recorded EEG signals. As data from each channel is a composite of multiple signals, de-mixing of the signal for removing artifacts is necessary \cite{pion-tonachiniICLabelAutomatedElectroencephalographic2019a}.

    Artifacts in EEG data can be categorized into two types: external and internal. External artifacts, including line noise, channel noise, and environmental noise, can be effectively removed due to their limited frequency or spatial constraints. Internal artifacts involve Electrooculography (EOG), Electrocardiography (ECG), and Electromyography (EMG) signals. EOG artifacts, related to eye movement, are concentrated on the frontal head surface, while ECG artifacts cover the entire head surface. The patterns of EOG and ECG signals are distinctive, regular, or periodic, often also accompanied by reference signals. Consequently, established methods like regression, adaptive filtering, statistical measures \cite{jasAutorejectAutomatedArtifact2017, islamProbabilityMappingBased2019, Islam2021}, or blind source separation \cite{pion-tonachiniICLabelAutomatedElectroencephalographic2019a,Shoker2005,Nolan2010, mognonADJUSTAutomaticEEG2011,Bigdely-Shamlo2013,winklerAutomaticClassificationArtifactual2011,winklerRobustArtifactualIndependent2014,frolichClassificationIndependentComponents2015,chaumonPracticalGuideSelection2015} have proven effective in identifying and removing these artifacts.
    
    The most intricate artifacts to address are electromyogenic (EMG) artifacts. They manifest themselves across various regions of the head surface depending on the responsible muscle groups and are hence spatially broadly distributed. Their waveforms and amplitudes differ depending on (1) muscle tissues, (2) degree of contraction, and (3) a subject's sex \cite{Chen2019}. The frequency of EMG artifacts completely overlaps with EEG signals from 0Hz to \textgreater 200Hz. Unlike the effective separation of EOG and ECG artifacts by Independent Component Analysis (ICA), the segregation of EMG artifacts into distinct independent components is exceptionally challenging. The reference signal for muscle artifacts is rarely available and it is impractical to cover all muscle groups involved. Consequently, regression or filtering methods cannot be used \cite{mannanIdentificationRemovalPhysiological2018}. EMG artifacts, bearing similarities to EEG signals, can significantly impair the accuracy of analyses \cite{Chen2019}, ranging from brain disorder diagnosis to motor imagery classification and the interpretation of nervous system functionality. Effective artifact cleaning has been demonstrated to greatly enhance the quality of results \cite{abdi-sargezehEEGArtifactRejection2021}, particularly in scenarios involving small sample sizes and a limited number of EEG channels. Hence, the effective elimination of such artifacts is paramount. 
    
    This study focuses on effective EMG artifact epoch detection, an optional but potentially simpler and more straightforward step preceding EMG artifact removal. The assumption here is that effective EMG artifact epoch detection provides sufficient information for subsequent artifact removal, assuming an appropriate removal approach and also allowing direct removal of the epochs with low signal noise ratio in case the artifacts cannot be effectively removed. Given the diverse origins and patterns of EMG artifacts, our study specifically addresses EMG artifacts relevant to resting-state EEG recordings.  
    
    For investigating the cleaning of EMG artifacts from EEG recordings, data are typically obtained either by recruiting subjects to perform specific EMG artifact tasks or by integrating existing datasets. Prior studies have shown considerable variability in both approaches, particularly in the former, and tend to rely on concept-oriented data collection with minimal justification provided for the selection of artifacts, task duration, or the number of task repetitions \cite{mucarquerImprovingEEGMuscle2020, liuStateDependentIVAModel2021, chenHybridMethodMuscle2021,barbanAnotherArtefactRejection2021,zhaoMultistepBlindSource2021}.
    
    Additionally, a wide range of subject numbers can be found across studies: four to 200 subjects in analytical approaches \cite{Bigdely-Shamlo2015,chenHybridMethodMuscle2021,barbanAnotherArtefactRejection2021,zhaoMultistepBlindSource2021}, eleven to 200 subjects in classical machine learning approaches \cite{jasAutorejectAutomatedArtifact2017,winklerAutomaticClassificationArtifactual2011,winklerRobustArtifactualIndependent2014,saiFullyAutomatedUnsupervised2021,saba-sadiyaUnsupervisedEEGArtifact2021}, and seven subjects up to 6000 recordings (not clear if they were from different subjects) in deep learning (DL) approaches\cite{zhangEEGdenoiseNetBenchmarkDataset2021,sunNovelEndtoend1DResCNN2020,saba-sadiyaUnsupervisedEEGArtifact2021,Hwaidi2021,Zhang2019}. Since the collection of human experimentation data is rather expensive owing to time-consuming subject recruitment, long-time recording of each subject, cautious pre- and post-preparation and the necessary on-site recording personal, data sets published to date are often characterized by small sample sizes, in this case concerning the number of subjects \cite{barbanAnotherArtefactRejection2021,chenHybridMethodMuscle2021,zhaoMultistepBlindSource2021,Zhang2019,Hwaidi2021,saba-sadiyaUnsupervisedEEGArtifact2021,sunNovelEndtoend1DResCNN2020,zhangEEGdenoiseNetBenchmarkDataset2021, saiFullyAutomatedUnsupervised2021}. 
    
    This work concentrates on optimizing data collection, aiming to minimize costs while preserving the artifact cleaning model performance. We address five key questions:
    \begin{enumerate}
        \item \label{question_artifact_type} Which types of artifacts should be considered?
        \item \label{question_repetitions} How many repetitions are necessary for each artifact task?
        \item \label{question_cover_others} Is training on certain artifacts sufficient to detect other types? 
        \item \label{question_generalization} Can a trained model generalize to unknown subjects? 
        \item \label{question_pretraining} Can a pre-trained model offer advantages, e.g., via transfer learning?
    \end{enumerate}
    
    To answer these questions, we have conducted an EEG experiment involving different types of isometric contractions (referred to as contractions) and continuous movements (referred to as movements) serving as artifact generators as well as a task without voluntary artifact generation. For our analysis, we train subject-specific and generalized DL models to detect the artifact epochs using three different DL architectures, which are suitable for the data and problem at hand.

    This work presents the following noteworthy contributions beyond existing studies:
    \begin{itemize}
        \item Data-Oriented Design: Building upon prior concept-oriented approaches \cite{mucarquerImprovingEEGMuscle2020, liuStateDependentIVAModel2021, chenHybridMethodMuscle2021,barbanAnotherArtefactRejection2021,zhaoMultistepBlindSource2021}, our work extends towards data-oriented design based on a more systematic and dynamic quantitative analysis for data collection, ensuring more cost-effectiveness and higher data quality.
        
        \item Innovative EMG Channel Derivation: Instead of relying on conventional EMG recording methods, we derive EMG channels directly with the EEG channel combinations, which can be utilized due to their positions closely resembling bipolar EMG channels. This approach substantially reduces the preparatory and post-processing time and cost for each recording.
    \end{itemize}

\section{Material and Methods}
    
    \subsection{Recording Setup} 
        Conventionally, EEG and EMG artifacts are recorded using separate EEG and EMG channels. This presents challenges, particularly when recording from multiple muscle groups. Positioning additional EMG channels is a time-consuming process, requiring the manual identification of the correct muscle groups for each individual. 
        
\begin{figure}[!htb] 
        \centering
        \includegraphics[width= 0.95\linewidth]{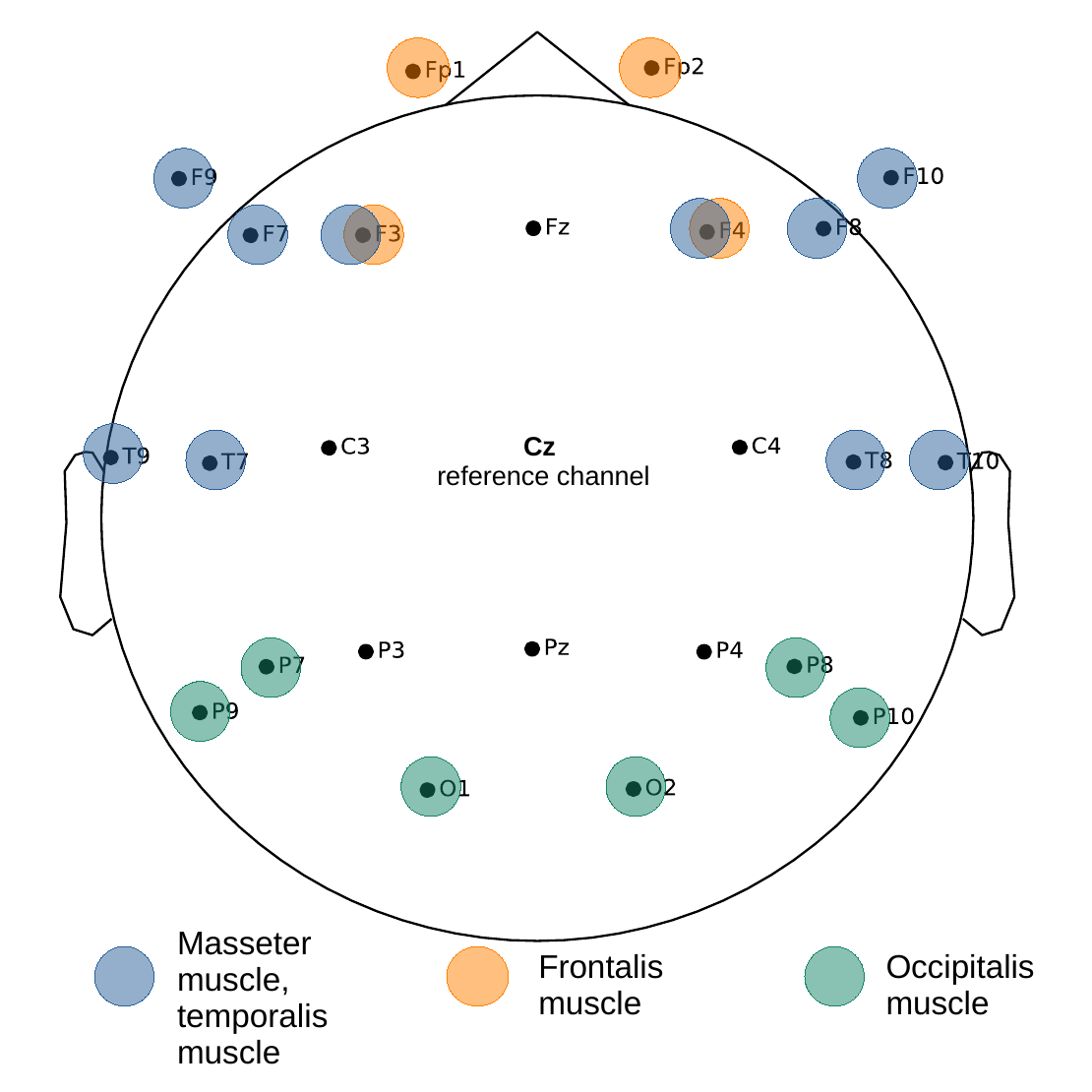} 
        \caption{EEG channel positions on the head. The marked channels are used to derive the according EMG signals.}
        \label{fig_EEG_channel_pos}
        \end{figure}

        In our work, we introduce an alternative solution, where we leverage the EEG channels taking advantages of their positions over the facial muscle groups of interest according to the EEG 10-10 system. These channels closely resemble bipolar EMG channels, allowing us to derive the corresponding EMG signals (\autoref{tab_channel_combi}, \autoref{fig_EEG_channel_pos}). The EEG caps used in our study are waveguard\texttrademark original (ANT Neuro GmbH, Berlin, Germany).
        
        The recordings in this study are acquired at a sampling rate of 2048Hz with 25 EEG channels (\autoref{fig_EEG_channel_pos}). 

        \begin{table}[!htb]
        \renewcommand{\arraystretch}{1.3}
        \caption{EEG channel combination for EMG channel derivation (also refer to \autoref{fig_EEG_channel_pos})}
        \label{tab_channel_combi}
        \centering
        \begin{tabularx}{\columnwidth}{|X|X|} 
        \hline
        \bfseries Bipolar Combinations & \bfseries Muscle Groups \\
        \hline\hline
        F9-F7, F9-F3, F7-F3, T9-T7, F9-T9, F9-T7, T9-F7, F10-F8, F10-F4, F8-F4, T10-T8, F10-T10, F10-T8, T10-F8 & Masseter muscle, Temporalis muscle \\
        \hline
        Fp1-F3, Fp1-Fp2, Fp1-F4, Fp2-F4, Fp2-F3 & Frontalis muscle \\
        \hline
        O1-P7, O1-P9, P7-P9, O1-O2, O2-P8, O2-P10, P8-P10 & Occipitalis muscle \\
        \hline
        \end{tabularx}
        \end{table}
        
    \subsection{Data Collection, Preprocessing and Feature Engineering} 
        
        Seven subjects participated in this study (age 20 – 47 years, five female and two male). All subjects were informed about the purpose and the methods of the study, and provided written consent. Two authors also participated in the experiments. Nonetheless, understanding the study's objectives does not interfere with the collected data. The study was approved by the Ethics Committee of the Bavarian State Medical Association, Germany (Bayerische Landesärztekammer, application number: 23035). 
        
        This study include twelve artifact types identified from the literature \cite{liElectromyogramEMGRemoval2021,barbanAnotherArtefactRejection2021,rantanenSurveyFeasibilitySurface2016} and commonly observed in daily clinical routine in the forms of isometric contractions and continuous movements (\autoref{tab_EMG_artifact_selection}). 
        
        \begin{table*}[!hbt] 
            \centering
            \caption{EMG artifact types}
            \label{tab_EMG_artifact_selection}
            \begin{tabularx}{\textwidth}{|c|X|X|X|} \hline 
                   & \bfseries Artifacts & \bfseries Source Muscle Group & \bfseries Artifact Type \\ \hline \hline 
                 1 & Jaw tensing & Masseter muscle, temporalis muscle & Isometric contractions \\ \hline 
                 2 & Biting & Masseter muscle, temporalis muscle & Continuous movements\\ \hline 
                 3 & Teeth grinding & Masseter muscle, temporalis muscle & Continuous movements\\ \hline 
                 4 & Frowning & Frontalis muscle & Isometric contractions\\ \hline 
                 5 & Eyebrows raising and holding & Frontalis muscle & Isometric contractions \\ \hline 
                 6 & Eyebrows up and down & Frontalis muscle & Continuous movements \\ \hline 
                 7 & Head turning left and holding & Occipitalis muscle & Isometric contractions\\ \hline 
                 8 & Head turning right and holding & Occipitalis muscle & Isometric contractions\\ \hline 
                 9 & Head turning left and right & Occipitalis muscle & Continuous movements\\ \hline 
                 10 & Head tilting downwards and holding & Occipitalis muscle & Isometric contractions \\ \hline 
                 11 & Head tilting upwards and holding & Occipitalis muscle & Isometric contractions \\ \hline 
                 12 & Nodding & Occipitalis muscle & Continuous movements \\\hline
            \end{tabularx}
        \end{table*}
        
        In order to keep the collected artifact data realistic, subjects were made comfortable, so that artifacts could be generated in a natural way. Based on feedback from subjects in a pilot study, we recorded isometric contraction epochs lasting 5 seconds and continuous movement epochs lasting 10 seconds. The number of repetitions for each artifact task was determined based on the maximum number that subjects felt comfortable with.
        
        Each subject participated in seven isometric contraction artifact tasks, each lasting five seconds and repeated ten times per task, and five continuous movement tasks lasting ten seconds and repeated five times per task. Consequently, there are a total of 95 artifact-containing epochs for each subject \footnote{One of the subjects has 94 artifact-containing epochs.}.
        
        For the non-artifact epochs, we utilize eyes-open (EO) resting-state recordings of 4.82$\pm$0.85 minutes for each subject. These recordings are segmented into alternating 10-second and 5-second epochs without overlap. In total, 38$\pm$7 EO epochs per subject are employed as non-artifact epochs in our analysis.

        The collected data are publicly available \cite{Wang-NoethEtAll2024BREAD}. 
        
        We use a high-pass filter with a cutoff frequency of 1Hz to eliminate low-frequency noise. Additionally, a notch filter is applied to attenuate the 50Hz power line interference and its harmonic frequencies up to 1001Hz. Both filtering techniques are implemented using MNE-Python \cite{GramfortEtAl2013a}. 
        
        It is important to note that no further pre-processing steps are carried out. This decision is based on our assumption that both the EO and the artifact epochs are inherently clean. Specifically, we refrain from manual artifact removal to prevent the introduction of any subjective bias into the training data. While there may be some artifacts present in the EO data, the variability in artifacts across different subjects is expected, and the collective use of this data is not anticipated to result in significant distortion of the EO signals.
               
\begin{figure} [!htb] 
        \centering
          \subfloat[]{\includegraphics[width=\linewidth]{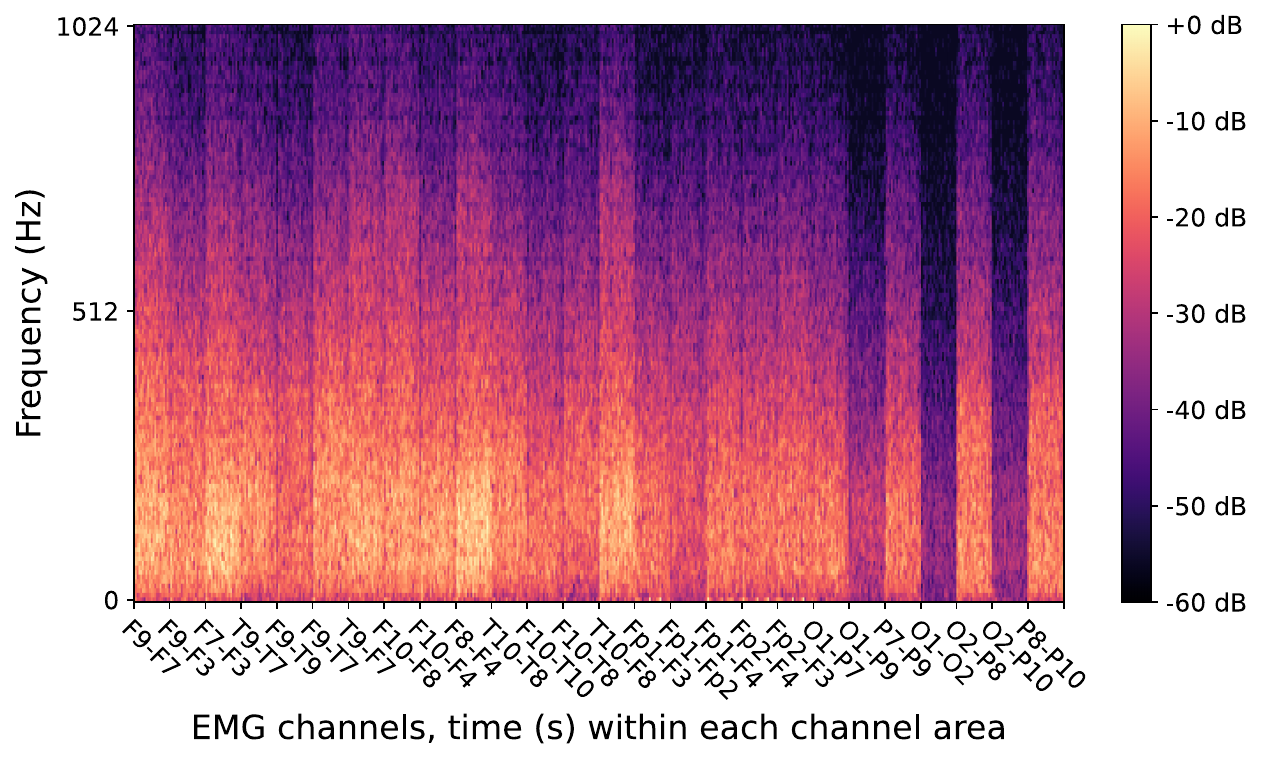}}\ 
          \subfloat[]{\includegraphics[width=\linewidth]{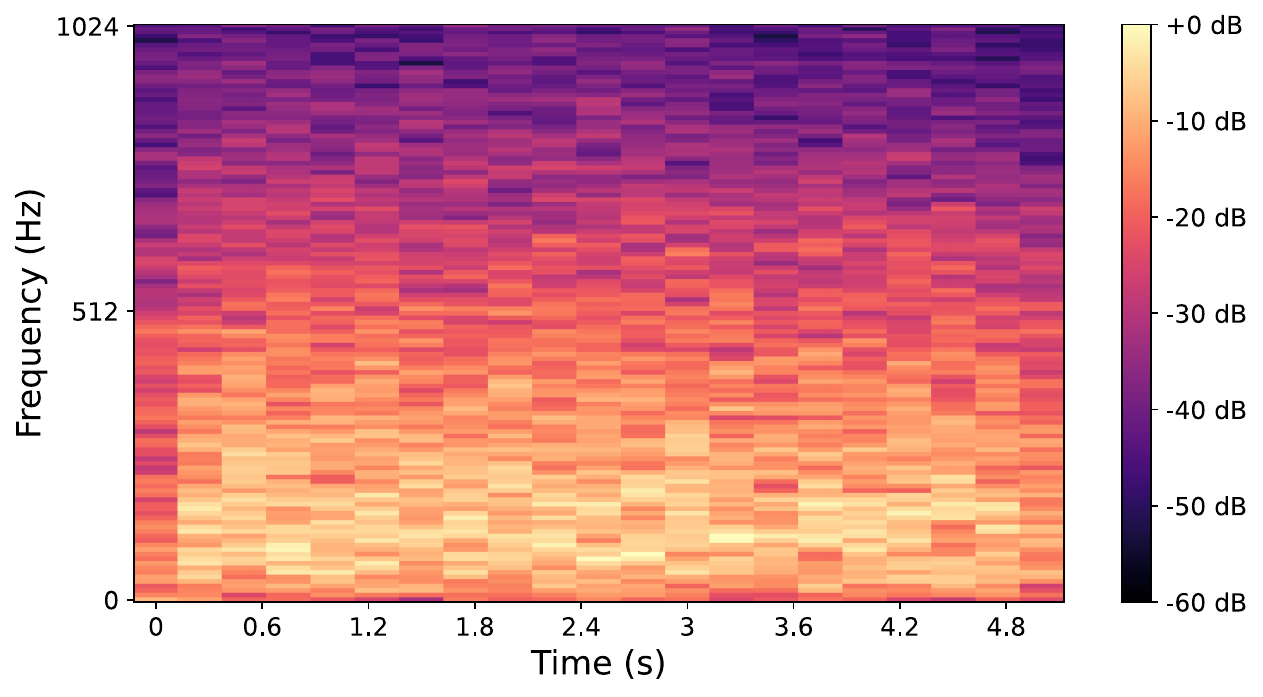}}\ 
          \subfloat[]{\includegraphics[width=\linewidth]{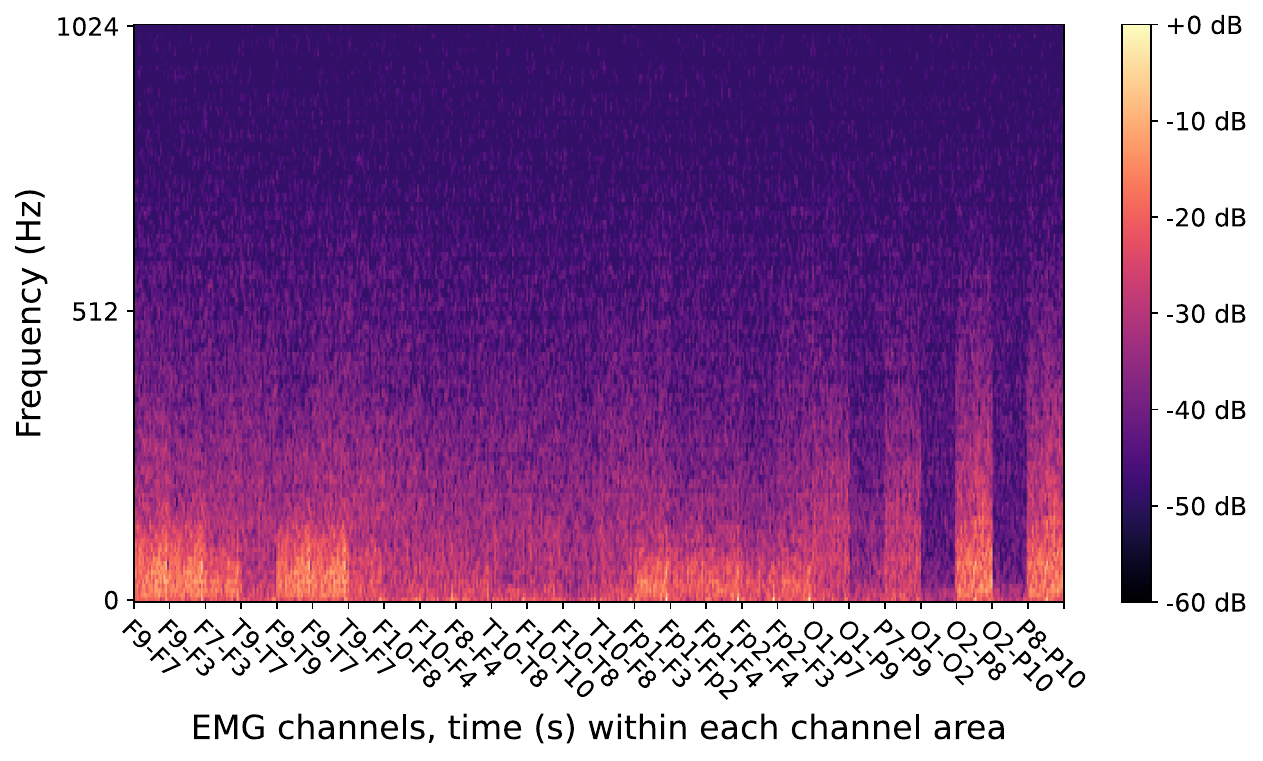}}
        \caption{(a) Illustration of a mel-spectrogram depicting an artifact epoch, particularly, the jaw tensing task. The EMG channels are concatenated along the x-axis, representing time within each channel area. (b) Zoomed-in view of channel F9-F7 from (a). (c) Exemplary mel-spectrogram showcasing a non-artifact epoch from EO recordings.}
        \label{fig_mel_spectro_kb_a_EO}
        \end{figure}

        We utilize a straightforward feature representation, namely spectrograms of each epoch. Spectrograms in mel scale are used, because a mel-spectrogram often has fewer dimensions than a traditional one due to frequency compression, reducing computational complexity and hence benefiting DL methods \cite{gaoEEGDrivingFatigue2023}. In the mel-spectrograms, time is represented along the horizontal axis and frequency along the vertical axis. We concatenate the EMG channels along the x-axis (\autoref{fig_mel_spectro_kb_a_EO}). The mel-spectrograms are generated using the Python library librosa 0.10.0\cite{mcfee2015librosa}, utilizing a Fast Fourier transform (FFT) window length of 409 (0.2 of the sampling rate) with a Hann window and no overlap. Power is calculated instead of energy.
       
     \subsection{Artifact Detection Models and Data Analysis}
        We have implemented a binary classification approach, distinguishing between EMG artifact epochs and non-artifact epochs based on mel-spectrograms. Three different CNN architectures including both classical and state-of-the-art ones have been tested: ResNet-18 \cite{heDeepResidualLearning2016}, Vision Transformer Small \cite{dosovitskiyImageWorth16x162021} and ConvNeXt Tiny \cite{liuConvNet2020s2022}. The selection of shallower versions for all three architectures has been made in consideration of our small sample size. The classification utilizes Python libraries fastai2 \cite{howard2020fastai} and timm 0.9.5 \cite{rw2019timm}, leveraging their built-in and pre-trained models resnet18, vit\_small\_patch32\_224 and convnext\_tiny. 

        The analysis has been structured into three phases (\autoref{tab_analysis_design}). First, we examine the impact of the number of repetitions and the generalizability across subjects within the artifact groups of isometric contractions and continuous movements (analysis \#1). Through this analysis, we can also assess the importance of different artifacts. Note that for isometric contractions, $7\times10$ epochs are employed per subject, and only $5\times5$ epochs for continuous movements. Epochs from non-artifact data are used in both analyses of isometric contractions and continuous movements. Next, we test for the possibility of generalization across artifact groups, i.e., training on isometric contractions and validation on continuous movements and vice versa (analysis \#2). Finally, we explore the approach of pre-training on the group and calibrating the individual models with the individuals' own data (analysis \#3). 
        
        All three analyses are additionally also tested for the possibility of reducing tasks by excluding isometric contractions of occipitalis muscle (artifact 7, 8, 10 and 11 in \autoref{tab_EMG_artifact_selection}). This exclusion is based on their relatively minor impact on EEG recordings, which will be further elaborated in the subsequent sections. Furthermore, they are more easily controlled compared to other isometric contraction artifacts such as jaw tensing, since subjects need to initiate strong head movements and maintain the position to generate these isometric contraction artifacts.
        
        \begin{table*}[!ht]
            \centering
            \caption{Analysis design}
            \label{tab_analysis_design}
            \begin{tabularx}{\textwidth}{|c|X|X|X|} 
            \hline
                  &\bfseries Analysis & \bfseries  Level & \bfseries Purpose \\ \hline \hline
                  
                  \#1& Separate binary classification for (1) isometric contractions and (2) continuous movements 
                  & Generalized models: subject cross-validation; \newline and individual models: one person, one model 
                  & To answer key questions \labelcref{question_artifact_type,question_repetitions,question_generalization} \\ \hline
                  
                  \#2& Separate binary classification with (1) isometric contractions as training data, and continuous movements as validation data, (2) vice versa
                  & Generalized models: subject cross-validation; \newline and individual models: one person, one model 
                  & To answer key questions \labelcref{question_cover_others,question_generalization} \\ \hline
                  
                  \#3&(1) Pre-training with calibration integrating analysis \#1; \newline
                      (2) Pre-training with calibration integrating analysis \#2
                  & Group-level pre-training with individual-level calibration
                  & To answer key question \labelcref{question_pretraining} \\ \hline
                  
            \end{tabularx}
        \end{table*}
        
        Due to the small amount of available data, the models are trained and validated using cross-validation. For individual models in analysis \#1, ten models for isometric contractions per subject undergo 10-fold cross-validation, where each repetition of every task is grouped into one fold and validated. Non-artifact epochs are divided into ten folds, paired with 10-fold artifact epochs. Performance is documented after training with each fold of the nine cumulatively. Each trained model from analysis \#1 is also validated on the other artifact type (analysis \#2). Validation data includes all data from the other artifact type combined with non-artifact epochs from the validation data of analysis \#1. For continuous movements, five models per subject undergo validation using each of the five repetitions (analysis \#1). Non-artifact epochs are split into five folds, and the other steps are analogous to isometric contractions. Generalization across subjects is validated with the generalized models. In both analyses \#1 and \#2, seven models each for isometric contractions and continuous movements are validated, respectively, leaving one subject out for validation. The other steps are similar to those given above. Analysis \#3 follows the same steps as individual models in analyses \#1 (analysis \#3-(1)) and \#2 (analysis \#3-(2)), with the difference that each model is pre-trained with all repetitions of the same artifact type as the training data and all non-artifact data from all other subjects.
                
        Models are trained with a batch size of 64 or the size of the training dataset in cases where the training dataset is smaller than 64. The learning rate is set to 0.005, and the number of epochs to 80. Owing to the small sample size, hyperparameter tuning and the selection of optimized models have not been included. Therefore, in cross-validation for analysis \#1 and analysis \#3-(1), no extra test data is reserved.

        Given the imbalance in our training data, we evaluate our results using two distinct metrics: recall (measuring the percentage of correctly identified artifacts) and specificity (measuring the percentage of correctly identified non-artifact data). We expect to observe enhancements in the overall performance (mean performance) and a decrease in variability (standard deviation of performance) across all models as the number of repetitions increases.

        We present overall results with mean and standard deviation, followed by detailed results shown either by violin plots featuring kernel density estimates (KDE) of the underlying distribution of the observations or by swarm plots to display all observations for fewer amount of observations. The violin plots utilize the bandwidth method of Scott without smoothing beyond the extremes of the observed data. In addition, the 10th, 25th, 50th, 75th and 90th percentiles are also shown with box-and-whisker plots, indicated as the lower whisker, the lower quartile, the median, the upper quartile and the upper whisker. 

        Statistical comparisons of results obtained by two different analyses or two different algorithms are performed using the Wilcoxon Signed-Rank Test. Correlation coefficients ($\tau$) between results and the number of task repetitions are calculated using the Kendall’s Rank Correlation. The statistical significance level was set to $p<0.05$. The values of statistical tests are rounded to three decimal places. These tests utilize the Python library SciPy \cite{2020SciPy-NMeth}. 
        
        It is important to note that both the Wilcoxon Signed-Rank Test and Kendall's Rank Correlation assume that the observations in each sample are independent and identically distributed. In the individual models, observations within each subject are not independent due to the 10-fold or 5-fold cross-validation. However, the observations from the seven different subjects are independent of each other. In the generalized models, observations are not independent due to the Leave-One-Out-Cross-Validation. Hence, there could be a higher type I error. In each Wilcoxon Signed-Rank Test, the training data and the validation (or test) data in the two analyses or algorithms compared are exactly or partially the same.
        
\section{Results}

    To test for the possibility of reducing tasks, two variations are investigated, where the isometric contractions of the occipitalis muscle (artifact 7, 8, 10, and 11 in \autoref{tab_EMG_artifact_selection}) are included (referred to as full set of tasks) (\autoref{result_with_occipitalis}) and excluded (referred to as selected set of tasks) (\autoref{result_without_occipitalis}), respectively.
    
    \subsection{Full Set of Tasks} \label{result_with_occipitalis}

        Using all repetitions and subjects, for isometric contractions, both individual and generalized models show a mean recall exceeding 0.85, with a standard deviation below 0.13 (\autoref{fig_result_sum}, \autoref{tab_result_sum_full} in Appendix \ref{Appendix_result_sum_tables}). However, the specificity demonstrates poorer performance. Individual models have a mean ranging from 0.88 to 0.9 and a standard deviation from 0.18 to 0.21, while generalized models exhibit a mean of 0.59 and a standard deviation of 0.35. 
        
        Concerning continuous movements, in contrast, the specificity yields superior values compared to the recall, averaging above 0.93 with a standard deviation below 0.16 across all analyses. The recall attains similar values, except in analyses \#2 and \#3-(2), where the mean ranges between 0.48 and 0.63, with a standard deviation between 0.12 and 0.19.     

        \begin{figure}[!htb] 
            \centering
            \subfloat[]{\includegraphics[width=\linewidth]{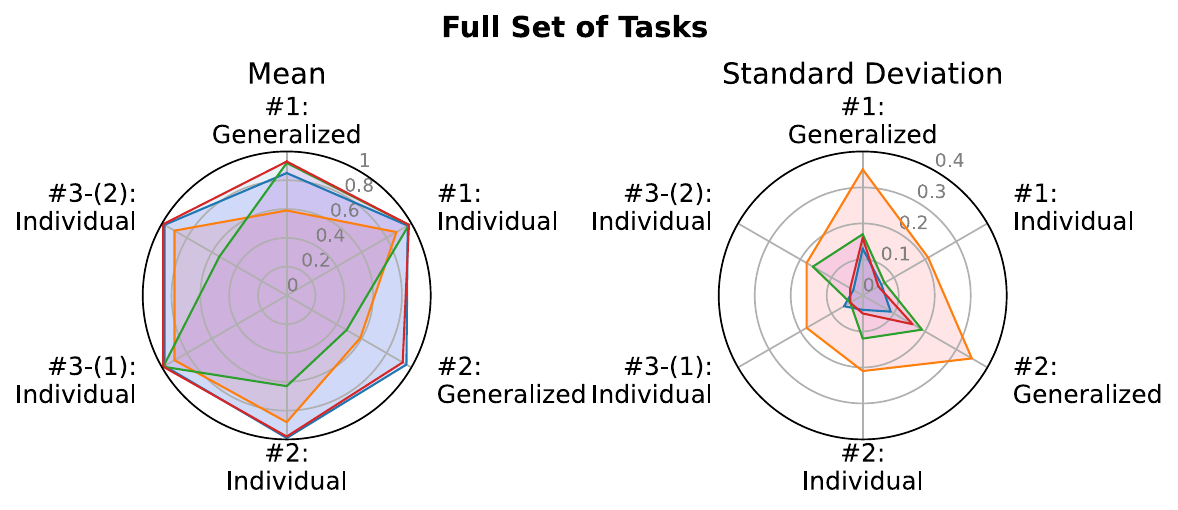}}\  
            \subfloat[]{\includegraphics[width=\linewidth]{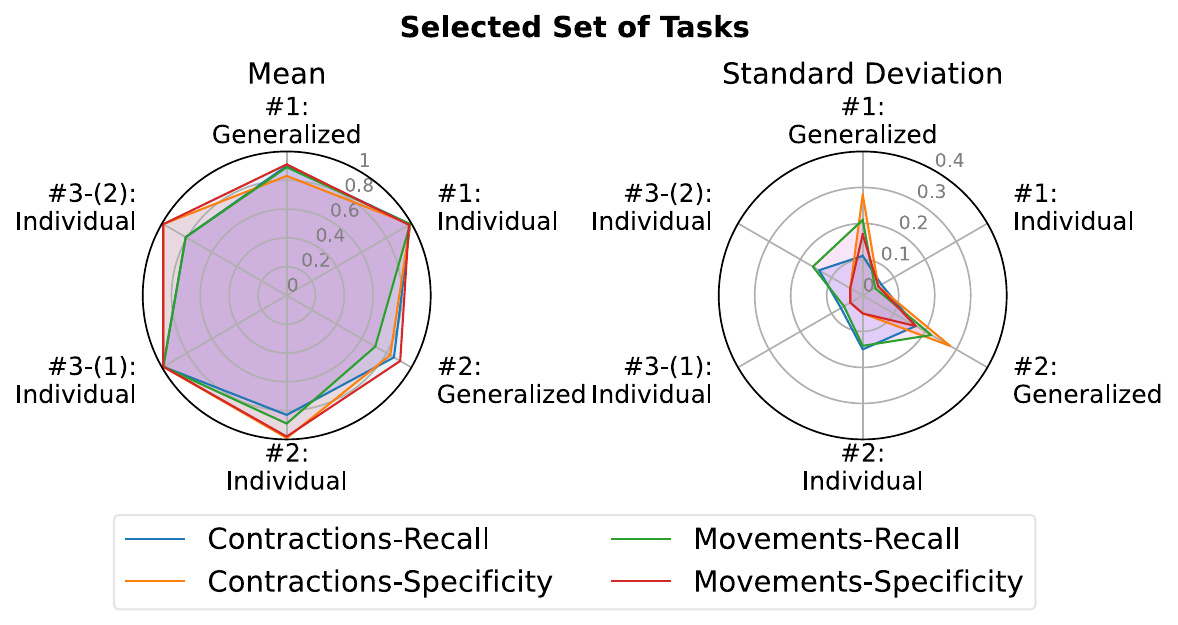}}\ 
            \caption{Overall results from all three analyses with models using all repetitions and all subjects in the form of the average of all three DL architectures (also refer to Tables \ref{tab_result_sum_full} and \ref{tab_result_sum_selected} in Appendix \ref{Appendix_result_sum_tables}).}
            \label{fig_result_sum}
        \end{figure}

        \subsubsection{Analysis \#1}

            \begin{figure*}[!bt]
                \centering
                \includegraphics[width=\linewidth]{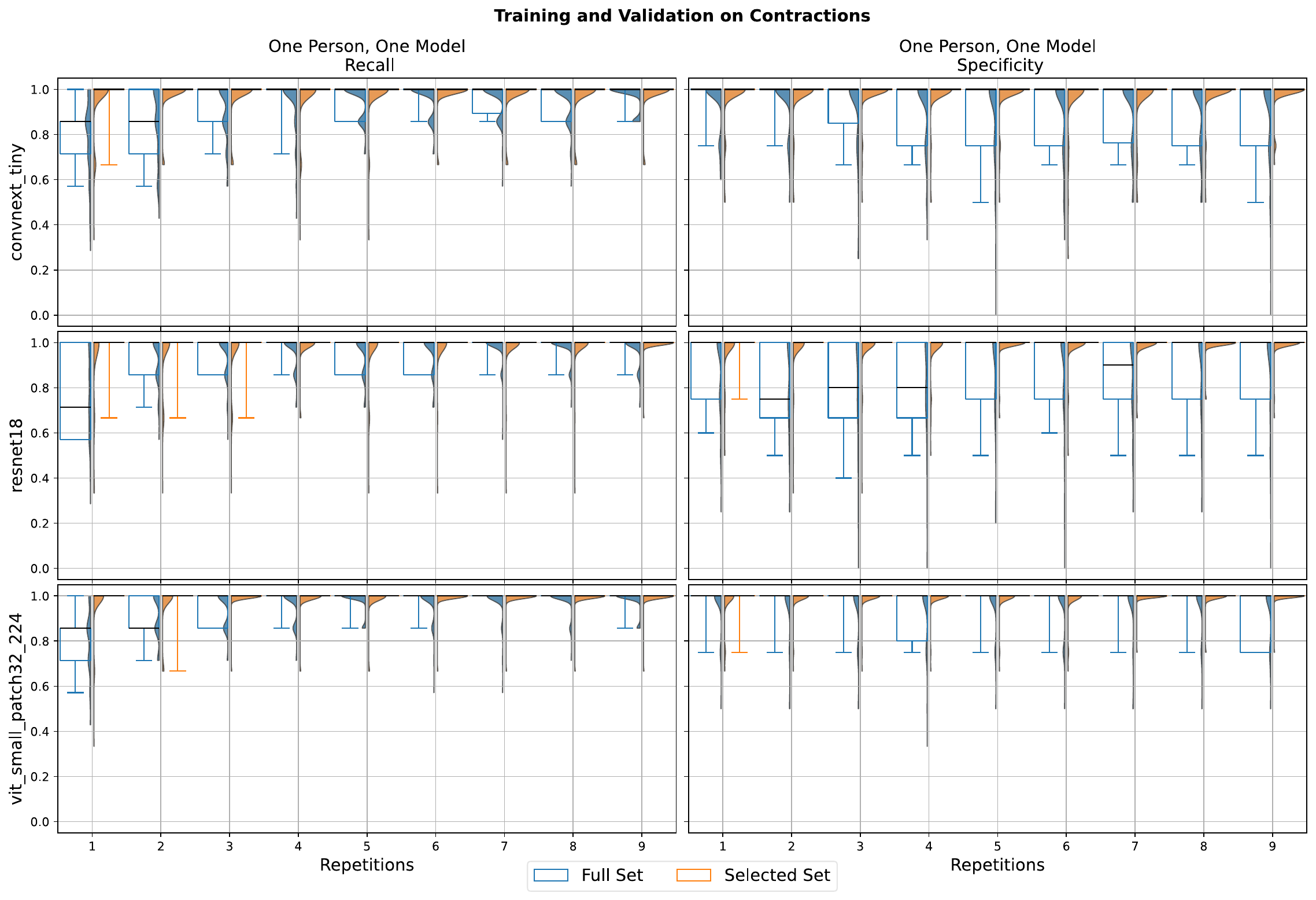}
                \caption{Analysis \#1: Performance of all three DL architectures for classifying isometric contraction artifacts with an increasing number of repetitions for one person, one model. The performance is shown with both the full and the selected set of tasks. For each repetition within a panel, there are 70 observations in both the full and the selected set, respectively. Each violin has the same area within a panel and whiskers in each box-and-whisker plot extend up to the 10th and the 90th percentiles. Some box-and-whisker plots are not shown because the lower whisker is close to one.}
                \label{fig_result_1_1}
            \end{figure*} 

            \begin{figure*}[!bt]
                \centering
                \includegraphics[width=\linewidth]{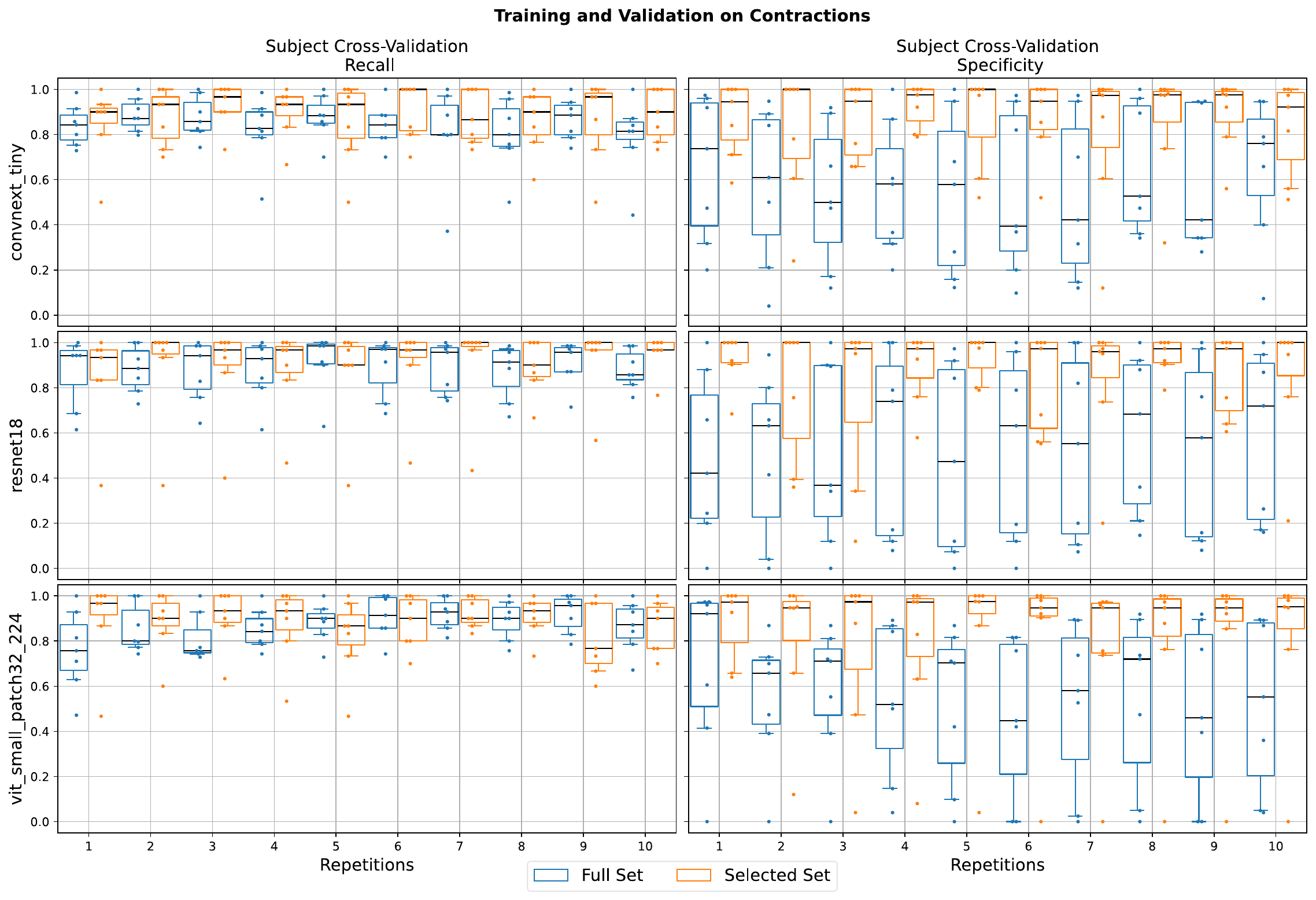}
                \caption{Analysis \#1: Performance of all three DL architectures for classifying isometric contraction artifacts with an increasing number of repetitions for subject cross-validation. Only the results trained with the maximal number of subjects (seven), are shown. The performance is shown with both the full and the selected set of tasks. For each repetition within a panel, there are seven observations in both the full and the selected set, respectively. Whiskers in each box-and-whisker plot extend up to the 10th and the 90th percentiles - in this case the second furthest observations, respectively.}
                \label{fig_result_1_2}
            \end{figure*}

            \begin{figure*}[!bt]
                \centering 
                \includegraphics[width=\linewidth]{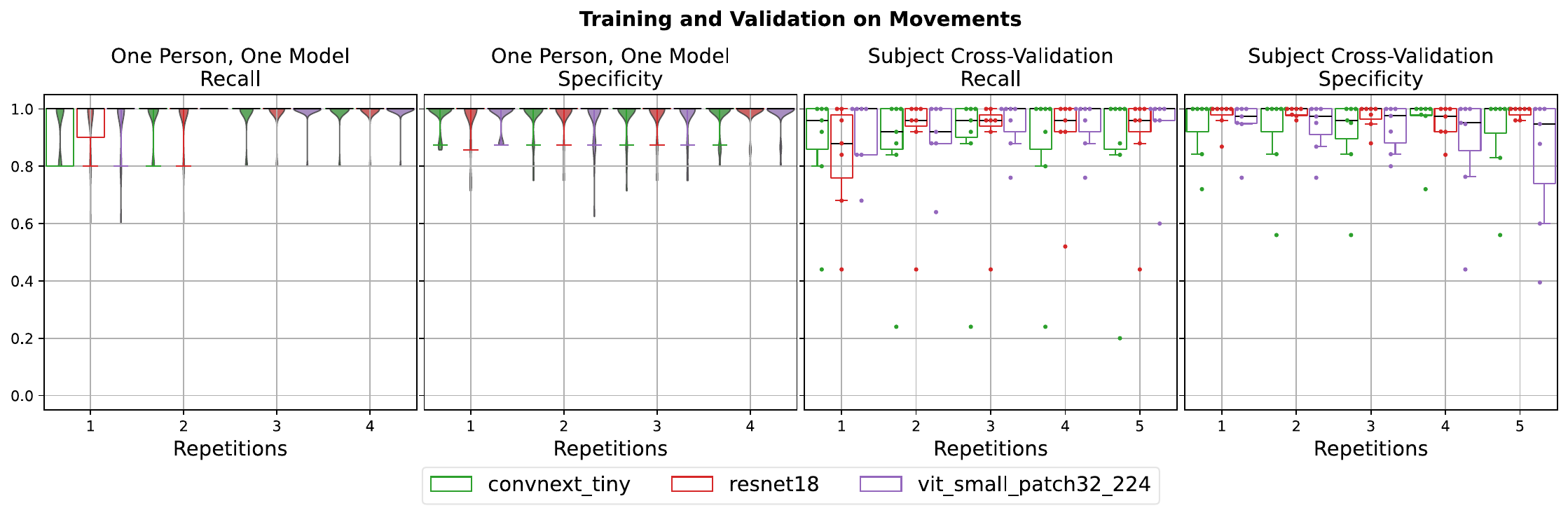}
                \caption{Analysis \#1: Performance of all three DL architectures for classifying continuous movement artifacts with an increasing number of repetitions for both experimental model types. Subject cross-validation only shows the results trained with the maximal number of subjects. The performance is not affected by excluding the four isometric contraction tasks. For each repetition and each architecture within a panel, there are 35 observations for one person, one model and seven for subject cross-validation, respectively. For one person, one model, each violin has the same area within a panel. Whiskers in each box-and-whisker plot extend up to the 10th and the 90th percentiles or the second furthest observations, respectively. Some box-and-whisker plots are not shown because the lower whisker is close to one. One violin plot is not shown because all observations are close to one.}
                \label{fig_result_1_3}
            \end{figure*} 
            
            For isometric contractions, while the recall improves with more repetitions in individual models ($\tau=0.330$, $p<0.0001$, \autoref{tab_statistic_test_rep_corr_full} in Appendix \ref{Appendix_statistical_tests}, \autoref{fig_result_1_1}), the specificity does not exhibit consistent improvement, especially for four subjects (\autoref{tab_analysis1_spec_t_1p1m} in Appendix \ref{Appendix_before_kaout}). Subjects vary notably in performance. In generalized models, no consistent improvements in both recall and specificity can be observed (\autoref{fig_result_1_2}, \autoref{tab_statistic_test_rep_corr_full} in Appendix \ref{Appendix_statistical_tests}). Due to the poor performance in specificity, these trained models cannot be applied directly onto a new subject.
            
            An investigation into misclassified items reveals that isometric contraction tasks involving head movements (head turning left, head turning right, looking up and looking down) produce EMG artifacts from the occipitalis muscle and are commonly misclassified as non-artifact epochs in both individual and generalized models (Figures \ref{fig_misclass_expt1_indi_before} and \ref{fig_misclass_expt1_group_before} in Appendix \ref{Appendix_before_kaout}). Further investigation has suggested that the mel-spectrograms of these artifact epochs indeed closely resemble those of non-artifact epochs among most subjects (\autoref{fig_mel_spectro_misclass} in Appendix \ref{Appendix_before_kaout}). However, it is unlikely that these artifacts are present in non-artifact epochs, as these movements are quite strong and would have been acknowledged and marked in the datasets during an EO session by the recording team conducting the data collection. 

            For continuous movements, in both individual and generalized models, there is no consistent improvement in both recall and specificity with an increasing number of repetitions (\autoref{fig_result_1_3}, \autoref{tab_statistic_test_rep_corr_full} in Appendix \ref{Appendix_statistical_tests}).

            Due to the instability and large standard deviations (\autoref{fig_result_sum}), these trained models cannot be applied directly onto a new subject. Note that the dataset for continuous movements remains unchanged, whether using the full or the selected set of tasks.   

        \subsubsection{Analysis \#2} 
            Recording only isometric contractions to differentiate both artifact types from non-artifact epochs initially appears feasible in individual models; however, it is challenged by the relatively large standard deviation of the specificity, reaching as high as 0.21 (\autoref{fig_result_sum},\autoref{tab_result_sum_full} in Appendix \ref{Appendix_result_sum_tables}). 
            
            On the other hand, relying solely on recording continuous movements proves insufficient to distinguish both artifact types from non-artifact epochs. A recall of $0.63\pm0.12$ in individual models reveals that, even with training data on continuous movements from a subject, the model cannot effectively differentiate isometric contraction artifacts from non-artifact data for the same subject.

            In generalized models, neither modality is sufficient, with a specificity of $0.59\pm0.35$ and a recall of $0.48\pm0.19$, respectively.

        \subsubsection{Analysis \#3}
            In both analysis \#3-(1) and \#3-(2), when compared to the individual models of analysis \#1 and \#2, respectively, pre-training with calibration does not exhibit performance improvement for isometric contractions or continuous movements, except for recall of continuous movements comparing analysis \#3-(1) and \#1 ($p=0.042$, \autoref{tab_statistic_test_pre_indi} in Appendix \ref{Appendix_statistical_tests}, \autoref{fig_result_sum}).   

    \subsection{Selected Set of Tasks} \label{result_without_occipitalis}
        
        Since subjects need to initiate strong head movements and maintain the position to generate these isometric contraction artifacts, they are more easily controlled compared to other isometric contraction artifacts, such as jaw tensing. Consequently, we have decided to exclude these artifacts and to additionally conduct all three analyses without these artifacts.

        Using all repetitions and subjects, for isometric contractions, both individual and generalized models exhibit a mean recall exceeding 0.81, with a standard deviation below 0.17 (\autoref{fig_result_sum}, \autoref{tab_result_sum_selected} in Appendix \ref{Appendix_result_sum_tables}). The specificity demonstrates notably better performance, with a mean of 0.99 and a standard deviation below 0.05, except for generalized models, where the mean is 0.83 with a standard deviation of 0.28. 
        
        Regarding continuous movements, the recall averages above 0.81 with a standard deviation below 0.16, except for generalized models, where the mean ranges between 0.71 and 0.89 with a standard deviation between 0.21 and 0.22. The specificity exhibits superior values, with a mean above 0.98 and a standard deviation below 0.05, except for generalized models, where the mean averages at 0.91 with a standard deviation of 0.17.

        \subsubsection{Analysis \#1}    

            The exclusion of the four isometric contraction tasks notably improves the performance of isometric contractions, particularly in terms of specificity, in both individual ($p=0.005$ for recall and $p<0.0001$ for specificity) and generalized ($p=0.053$ for recall and $p<0.0001$ for specificity) models (Figures \ref{fig_result_1_1} and \ref{fig_result_1_2}). For continuous movements, there is no notable change in performance before and after excluding the four isometric contraction tasks, as the dataset for continuous movements remains unaffected. 
            
            Increasing the number of repetitions does not consistently enhance the performance of both isometric contractions and continuous movements in one person, one model (\autoref{tab_statistic_test_rep_corr_selected} in Appendix \ref{Appendix_statistical_tests}). For isometric contractions, ResNet-18 and Vision Transformer Small show a low correlation in both recall ($\tau=0.146$, $p<0.001$, on average) and specificity ($\tau=0.086$, $p=0.013$, on average), while ConvNeXt Tiny shows no correlation. Regarding continuous movements, all three architectures show a low correlation in recall ($\tau=0.203$, $p=0.008$, on average), however, no correlation in specificity. For isometric contractions, the lower whisker for both recall and specificity consistently hovers near one for most repetition counts across all three architectures (\autoref{fig_result_1_1}). Similarly, for continuous movements, the lower whisker for both recall and specificity remains above 0.8 across all repetitions and architectures (\autoref{fig_result_1_3}). This suggests a strong similarity between repetitions for the same subjects, allowing for a reduction in repetitions without noteworthy loss of information.   

            For the generalized models of both isometric contractions and continuous movements, an increase in the number of repetitions does not enhance the performance (\autoref{tab_statistic_test_rep_corr_selected} in Appendix \ref{Appendix_statistical_tests}). While the individual models consistently demonstrate a performance with concentrated distributions across all three architectures, the generalized models exhibit less stability and satisfaction with much broader distributions (Figures \ref{fig_result_1_2} and \ref{fig_result_1_3}). Especially for isometric contractions, even with the maximum number of subjects and repetitions, the lower quartile of the specificity for generalized models ranges roughly between 0.7 and 0.85, with the lower whisker ranging roughly between 0.55 and 0.75. It is worth noting that in subject cross-validation, only one model per repetition per architecture is trained for each subject, resulting in a total of of seven observations. This implies a less comprehensive sample for evaluating the performance statistics generalized models. In one person, one model, on the other hand, there are ten models for isometric contractions and five for continuous movements per repetition per architecture for each subject, resulting in a total of 70 and 35 observations, respectively.
            
        \subsubsection{Analysis \#2}           

            \begin{figure*}[!bt]
                \centering
                \includegraphics[width=\linewidth]{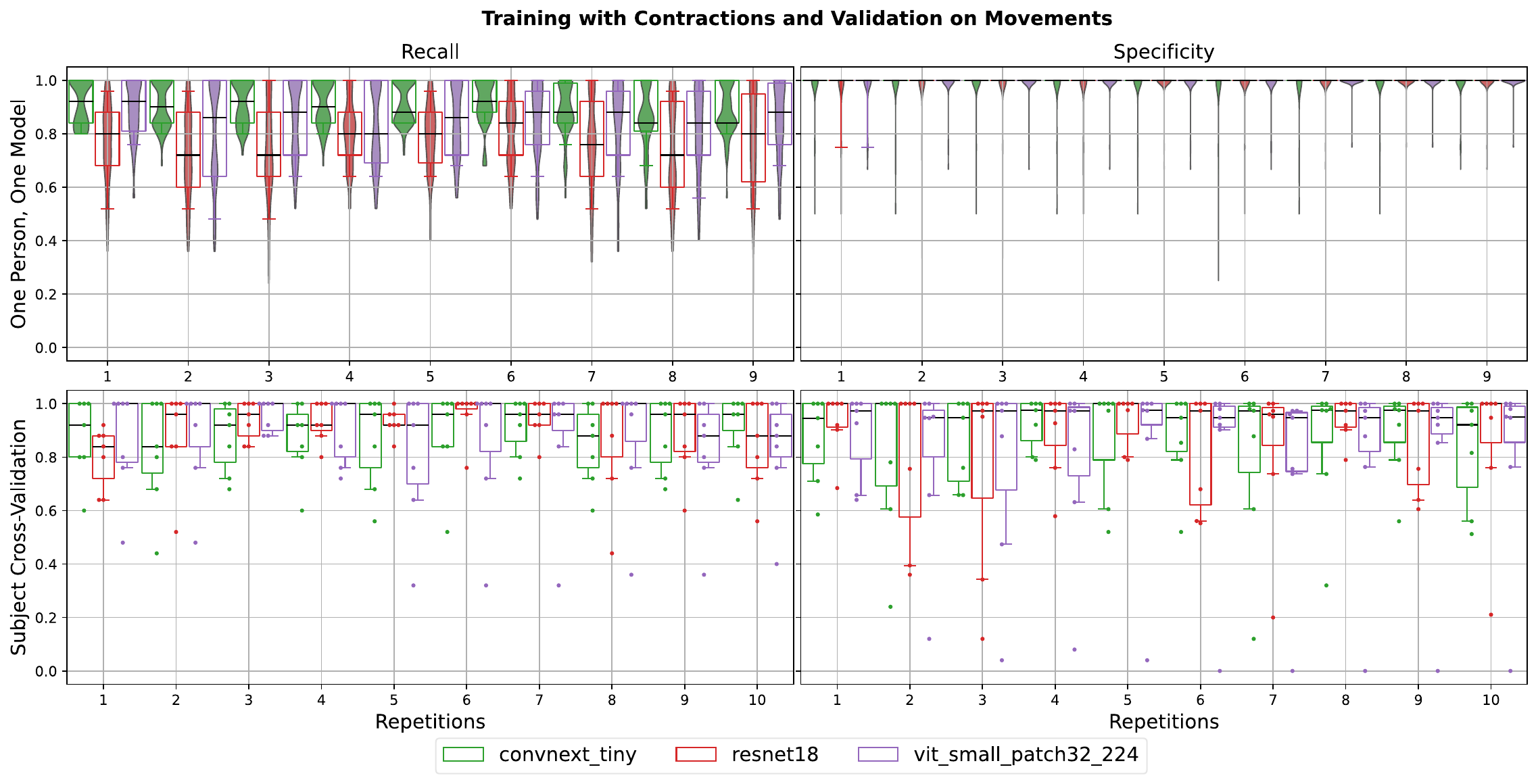}
                \caption{Analysis \#2: Performance of all three DL architectures for training with isometric contractions and validation on continuous movements with an increasing number of repetitions for both experimental model types. Subject cross-validation only shows the results trained with the maximal number of subjects. The performance is shown exclusively with the selected set of tasks. For each repetition and each architecture within a panel, there are 70 observations for one person, one model and seven for subject cross-validation. For one person, one model, each violin has the same area within a panel. Whiskers in each box-and-whisker plot extend up to the 10th and the 90th percentiles or the second furthest observations, respectively. Some box-and-whisker plots are not shown because the lower whisker is close to one.}
                \label{fig_result_2_tension}
            \end{figure*} 

\begin{figure*}[!bt]
                \centering
                \includegraphics[width=\linewidth]{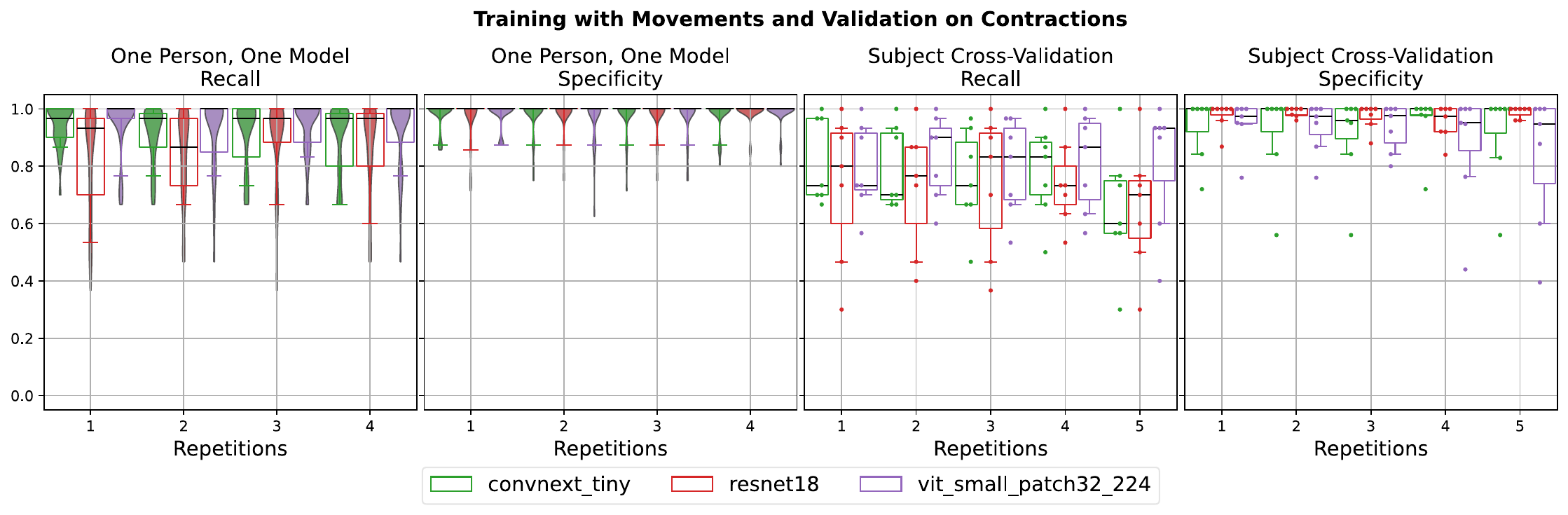}
                \caption{Analysis \#2: Performance of all three DL architectures for training with continuous movements and validation on isometric contractions with an increasing number of repetitions for both experimental model types. Subject cross-validation only shows the results trained with the maximal number of subjects. The performance is shown exclusively with the selected set of tasks. For each repetition and each architecture within a panel, there are 35 observations for one person, one model and seven for subject cross-validation. For one person, one model, each violin has the same area within a panel. Whiskers in each box-and-whisker plot extend up to the 10th and the 90th percentiles or the second furthest observations, respectively. Some box-and-whisker plots are not shown because the lower whisker is close to one.}
                \label{fig_result_2_movements}
            \end{figure*}

            In one person, one model, recording either only isometric contractions or continuous movements to differentiate both types from non-artifact epochs is feasible after excluding the four isometric contraction tasks. 
            According to the statistical test (\autoref{tab_statistic_test_best_result} in Appendix \ref{Appendix_statistical_tests}), there is no absolute best architecture for training with isometric contractions and validation on continuous movements or vice versa. However, for the former, using the ConvNext Tiny models consistently yield stable distributions (\autoref{fig_result_2_tension}). The median recall remains above approximately 0.85 for all repetitions, with a lower whisker above 0.8 for most repetitions, and the values for the specificity approach 1 for most repetitions. 
            For the latter, the results obtained with the Vision Transformer Small models also show a stable distribution with a median recall near 1 and a lower whisker above roughly 0.75 across all repetitions (\autoref{fig_result_2_movements}).
            This suggests a strong similarity between isometric contractions and continuous movements for the same subjects. More repetitions do not yield a considerable performance improvement (\autoref{tab_statistic_test_rep_corr_selected} in Appendix \ref{Appendix_statistical_tests}).

            Training on continuous movements and validation on isometric contractions exhibits a slightly better recall for one person, one model ($0.89\pm0.14$, \autoref{tab_result_sum_selected} in Appendix \ref{Appendix_result_sum_tables}) than vice versa ($0.83\pm0.15$), primarily because two continuous movement tasks targeting the occipitalis muscle (head turning left and right, and nodding) have not been excluded along with the isometric contraction tasks of this muscle group. These two tasks are the most frequently misclassified epochs (\autoref{fig_misclass_expt2_indi_after} in Appendix \ref{Appendix_after_kaout}). 
            
            For subject cross-validation, although the median for both recall and specificity for training with isometric contractions and validation on continuous movements exceeds 0.85 for all repetitions across all three architectures, the interquartile range (IQR) and the specificity distribution show considerable instability, with a notably broad spread (\autoref{fig_result_2_tension}). Despite a specificity of $0.91$ using all repetitions (\autoref{tab_result_sum_selected} in Appendix \ref{Appendix_result_sum_tables}), the recall for training with continuous movements and validation on isometric contractions drops as low as $0.71$, resulting in a poorer outcome compared to vice versa, where the balanced accuracy is 0.81 and 0.845, respectively.
            
        \subsubsection{Analysis \#3}
            
\begin{figure*}[!bt] 
                \includegraphics[width=\linewidth]{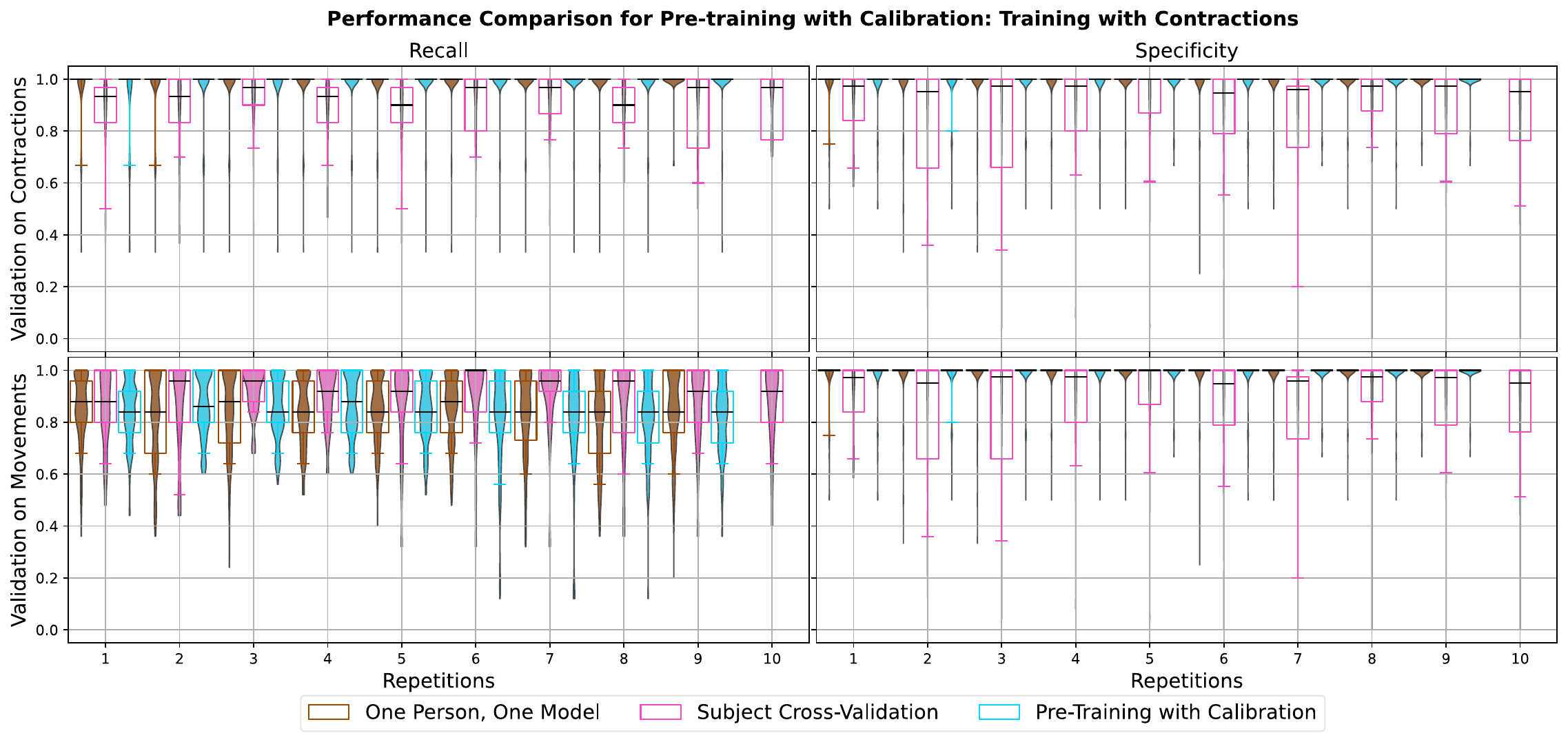} 
                \caption{Performance comparison between pre-training with calibration and the other two experimental model types for training with isometric contractions and validation on both isometric contractions and continuous movements with an increasing number of repetitions. The first row compares the results between analysis \#1 and analysis \#3-(1), while the second row compares analysis \#2 and analysis \#3-(2). Subject cross-validation only shows the results trained with the maximal number of subjects. The performance is shown exclusively with the selected set of tasks and the results for all three architectures are plotted together. For each repetition within a panel, there are 210 observations for one person, one model, 210 for pre-training with calibration and 21 for subject cross-validation. The x-axis has one less value for one person, one model and pre-training with calibration, because task repetitions were cross-validated keeping one repetition in the validation data. Each violin has the same area within a panel. Whiskers in each box-and-whisker plot extend up to the 10th and the 90th percentiles. Some box-and-whisker plots are not shown because the lower whisker is close to one.}
                \label{fig_result_3_tension}
            \end{figure*}

\begin{figure*}[!bt]
                \includegraphics[width=\linewidth]{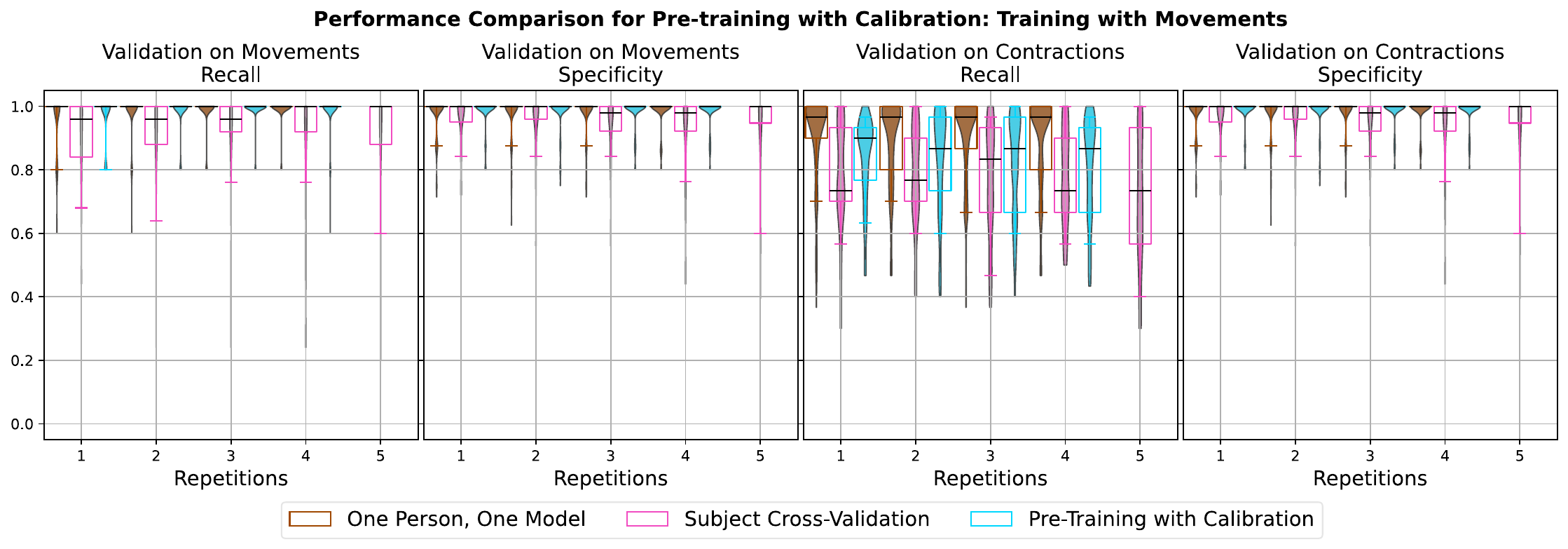}
                \caption{Performance comparison between pre-training with calibration and the other two experimental model types for training with continuous movements and validation on both isometric contractions and continuous movements with an increasing number of repetitions. The first two columns compare the results between analysis \#1 and analysis \#3-(1), while the last two columns compare analysis \#2 and analysis \#3-(2). Subject cross-validation only shows the results trained with the maximal number of subjects. The performance is shown exclusively with the selected set of tasks and the results for all three architectures are plotted together. For each repetition within each panel, there are 105 observations for one person, one model, 105 for pre-training with calibration and 21 for subject cross-validation. The x-axis has one less value for one person, one model and pre-training with calibration, because task repetitions were cross-validated keeping one repetition in the validation data. Each violin has the same area within a panel. Whiskers in each box-and-whisker plot extend up to the 10th and the 90th percentiles. Some box-and-whisker plots are not shown because the lower whisker is close to one.}
                \label{fig_result_3_movements}
            \end{figure*}

            Pre-training does not offer advantages compared to one person, one model (Figures \labelcref{fig_result_sum,fig_result_3_tension,fig_result_3_movements} and \autoref{tab_statistic_test_pre_indi} in Appendix \ref{Appendix_statistical_tests}).

\section{Discussion and Prospects for Future Research}

    In previous studies examining EMG artifact cleaning, data collection has typically involved selection of artifact tasks and determining the number of task repetitions with limited justification \cite{mucarquerImprovingEEGMuscle2020, liuStateDependentIVAModel2021, chenHybridMethodMuscle2021,barbanAnotherArtefactRejection2021,zhaoMultistepBlindSource2021}. To the best of our knowledge, no study yet has explored the optimal and minimal selection of EMG artifact tasks as well as the minimal task repetitions. Therefore, this study represents the initial efforts in this specific domain, which should make it a valuable contribution to the field. 
    
    In this work, we have achieved to minimize the data collection effort while maintaining satisfactory model performance trained with these data. By excluding less relevant artifacts and leveraging the strong similarity among repetitions of the same tasks and between isometric contractions and continuous movements of the same subjects, we have optimized our approach.

    With the full set of tasks, the performance of the individual models trained and validated on isometric contractions exhibits a relatively large standard deviation, which poses a hindrance when it comes to application. In contrast, the individual models trained with the selected set of tasks can be utilized without hesitation. The individual models trained and validated on continuous movements consistently exhibit reliable performance. The individual models trained with the full set of tasks do not facilitate training on isometric contractions and validation on continuous movements, nor the reverse. In contrast, with the selected set of tasks, both modalities are feasible. The generalized models trained with both the full and the selected sets in all analyses are not recommended for application. With both the full and the selected sets of tasks, pre-training with calibration yields similar performance as one person, one model.
    
    Regarding the scope of artifacts (key question \ref{question_artifact_type}), we have identified four isometric contraction artifacts of the occipitalis muscle as less relevant. Due to their easier controllability during recording compared to other artifacts, they can be excluded in future research. Individual models in analysis \#1 reveal that these four artifacts are often misclassified, since their mel-spectrograms closely resemble those of non-artifact epochs. While theoretically possible, the presence of the four artifacts in the non-artifact data, specifically in this case the EO resting-state data, is unlikely. This is because they can be easily recognized based on clearly visible head movements and marked in the datasets. The more plausible explanation is that these artifacts have a weaker impact on EEG recordings, because the muscle group is relatively distant from the scalp compared to other muscle groups. 

    After excluding the four isometric contraction artifacts, with an increasing number of repetitions, no remarkable improvement of individual models in analyses \#1 and \#2 has been observed. Overall, the performance with different numbers of repetitions remains stable, indicating the high similarity of repetitions of the same subjects. This allows us to reduce the repetitions to a large extent, i.e., one (or three for training, validation, and testing if necessary) (key question \ref{question_repetitions}).

    With the selected set of tasks, the results of the individual models in analysis \#2 indicate that recording only either isometric contractions or continuous movements to distinguish both artifact types from non-artifact epochs is viable (key question \ref{question_cover_others}). Training with continuous movements has shown a slightly better performance, as two continuous movement artifacts of the occipitalis muscle remain in the validation data. In experiments without continuous movements of occipitalis, the performance of training on either of the two artifact types is better comparable. Considering that continuous movements are more discernible to the recording team and controllable by the subjects, it is expected that isometric contraction tasks occur more frequently and, therefore, should receive more attention. Hence, our preferred approach is to record isometric contractions to address both artifact types. 

    In all three analyses, results for subject cross-validation are unstable and generalized models are mostly not suitable when seeking to directly apply a trained model to a novel subject. This suggests that the sample size is insufficient for generalization (key question \ref{question_generalization}). Future studies should explore a larger subject pool to determine the minimal sample size necessary for generalization.

    Owing to the limited sample size and the strong similarity of artifacts from the same subjects within the current dataset, pre-training does not seem to provide substantial advantages compared to one person, one model (key question \ref{question_pretraining}). Nevertheless, we anticipate that advantages such as higher and more stable performance may be possible as the sample size increases and if the degree of artifact correlation across subjects also increases. 
        
    As a result of all analyses, we have successfully reduced the number of artifact tasks from twelve to just three by eliminating four isometric contraction tasks and leveraging isometric contraction tasks for training to identify both artifact types. The number of repetitions for these isometric contraction tasks has been scaled down from ten to a minimum, i.e., one (or three for training, validation, and testing if necessary).
        
    Our study operates under the assumption that robust EMG artifact epoch detection equates to providing sufficient information for effective EMG artifact removal. Nonetheless, it is important to acknowledge that as more analysis steps are introduced, some information loss is inevitable, impacting the final performance. Recognizing and mitigating this information loss remains an open challenge that warrants further exploration.
    
    A noteworthy limitation of our study is that each artifact epoch spanned the entire duration of the recording without artifact-free intervals. The epochs have fixed durations of approximately 5 or 10 seconds. In future investigations, it will be valuable to explore the optimal segment length for training data. This will require comprehensive statistics on the typical duration of artifacts, which can be instrumental in refining artifact identification and removal processes. 

    Another limitation is that non-artifact epochs in the training data include only EO recordings. Direct application of the trained model to recordings with other mental states than the resting-state could potentially lead to the misidentification of brain signals from more intense mental activities as artifacts. Caution is advised in such scenarios.
           
    As for the features employed in our study, we have exclusively utilized mel-spectrograms. To unlock deeper insights and additional information, future work could explore the integration of spatial and functional connectivity data. The incorporation of such information, possibly through graph neural networks (GNNs) \cite{linFatigueDrivingRecognition2021,changClassificationFirstEpisodeSchizophrenia2021,demirEEGGNNGraphNeural2021,waghEEGGCNNAugmentingElectroencephalogrambased2020}, promises to enrich the analysis and improve artifact detection and removal accuracy. Additionally, it is important to note that while mel-spectrograms can accelerate computation, they may lead to information loss during frequency compression. Future research should compare the performance and computational speed of spectrograms with and without mel-scaling to achieve an optimal trade-off.

    Our models, at the current stage, are not subjected to hyperparameter tuning, primarily due to the limited sample size. In the future, with an expanded dataset, hyperparameter optimization offers the potential for performance enhancement. Fine-tuning the models and optimizing key parameters will be crucial in maximizing their efficiency.
    
    The quest of optimizing EMG artifact data collection and processing is ongoing, and our study lays the foundation for further developments in this vital domain of research. Future work can explore innovative techniques, larger and more divers datasets, and advanced machine learning methodologies, striving to optimize data collection and enhance artifact detection and removal methods. These advancements hold the potential to considerably benefit the broader scientific community working with EEG and EMG data.

\section{Conclusion}
    Given the substantial cost associated with collecting EEG data, this study establishes an efficient data collection framework that not only minimizes the data acquisition effort but also upholds the performance of models trained on this data. 
    
    The data collection design has been successfully optimized by implementing two key strategies:
    \begin{itemize}
        \item \textbf{Artifact Task Optimization:} The number of artifact tasks has been substantially reduced from an initial twelve to a more focused set of three -- jaw tensing, frowning as well as eyebrows raising and holding. This optimization stems from eliminating the recording of all isometric contraction tasks involving the occipitalis muscle as well as all continuous movement tasks.
                 
        \item \textbf{Repetition Reduction:} The number of repetitions for each isometric contraction epoch has been significantly cut down from an original ten to a minimal and more efficient quantity, often as low as three or even just one repetition. This reduction not only preserves resources, but maintains comparable results, rendering the data collection process more economical.
    \end{itemize}
    
    In summary, this research advances the methodology of data collection in the context of EEG and EMG studies, paving the way for more efficient, economical and data-oriented approaches. These improvements not only reduce the financial burden of data acquisition, but also enhance the quality and effectiveness of subsequent analyses, making this work a valuable contribution to the scientific community.
    
\section*{Acknowledgment}

    This work was partially funded by Federal Ministry for Economic Affairs and Climate Action of Germany (grant number ZIM KK5211501BM0). The research was conducted in accordance with the principles embodied in the Declaration of Helsinki and in accordance with local statutory requirements. All participants gave written informed consent to participate in the study. The findings may contribute to the development of products, the right of use to which is licensed to brainboost GmbH.
    
    Author contributions are as following. Concept: Lu Wang-Nöth, Hai Huang, Alexandra Reichenbach and Helmut Mayer. Methodology: Lu Wang-Nöth, Philipp Heiler, Hai Huang and Alexandra Reichenbach. Experiments: Philipp Heiler, Daniel Lichtenstern, Luis Flacke and Linus Maisch. Data curation: Lu Wang-Nöth, Daniel Lichtenstern, Luis Flacke and Linus Maisch. Formal analysis and software: Lu Wang-Nöth. Supervision: Hai Huang and Helmut Mayer. Writing – original draft: Lu Wang-Nöth. Writing – review and editing: all authors.

\bibliographystyle{IEEEtran}
\bibliography{MyLibrary,other_refs}

\clearpage
\newpage
\appendix
\setcounter{figure}{0}
\setcounter{table}{0}
\counterwithin{figure}{section} 
\counterwithin{table}{section}

    \subsection{Overall Results from All Three Analyses with Models Using All Repetitions and All Subjects} \label{Appendix_result_sum_tables}
    
        The overall results are presented in the form of the average of all three DL architectures with the full (\autoref{tab_result_sum_full}) and the selected set of tasks (\autoref{tab_result_sum_selected}), respectively.
    
        \begin{table} [!hbt]
            \centering
            \caption{Overall results with the full set of tasks from all three analyses with models using all repetitions and all subjects in the form of the average of all three DL architectures.}
            \label{tab_result_sum_full}
            \includegraphics[width=0.8\linewidth]{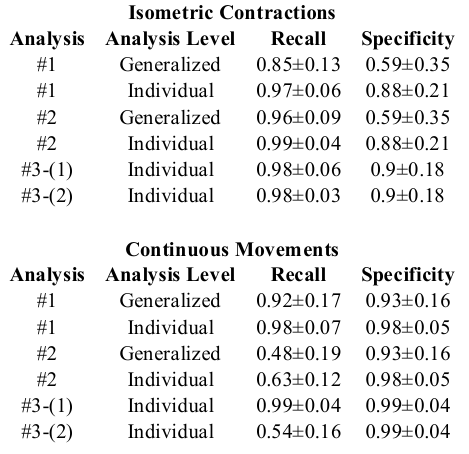} 
        \end{table}
        
        \begin{table} [!hbt]
            \centering
            \caption{Overall results with the selected set of tasks from all three analyses with models using all repetitions and all subjects in the form of the average of all three DL architectures.}
            \label{tab_result_sum_selected}
            \includegraphics[width=0.8\linewidth]{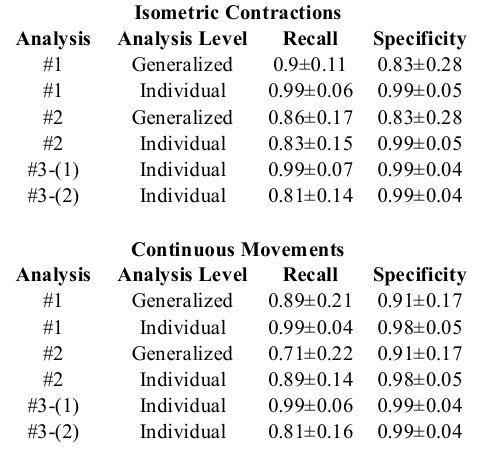} 
        \end{table}

    \subsection{Statistical Tests} \label{Appendix_statistical_tests}
        Statistical tests are performed to determine, 
        \begin{enumerate}
            \item if there is correlation between the number of repetitions and performance with the full (\autoref{tab_statistic_test_rep_corr_full}) and the selected set of tasks (\autoref{tab_statistic_test_rep_corr_selected}), respectively;
            \item if pre-training combined with calibration improves individual models (\autoref{tab_statistic_test_pre_indi});
            \item if an architecture yields the best result in one person, one model for analysis \#2 with the selected set of tasks (\autoref{tab_statistic_test_best_result}).
        \end{enumerate}
        Values are rounded to three decimal places, with 0.000 indicating values smaller than 0.0005.
        
        \begin{table*} [!hbt]
            \centering
            \caption{Correlation coefficient and $p$-value of statistical tests measuring the correlation between the number of repetitions and performance with the full set of tasks.}
            \label{tab_statistic_test_rep_corr_full}
            \includegraphics[width=\linewidth]{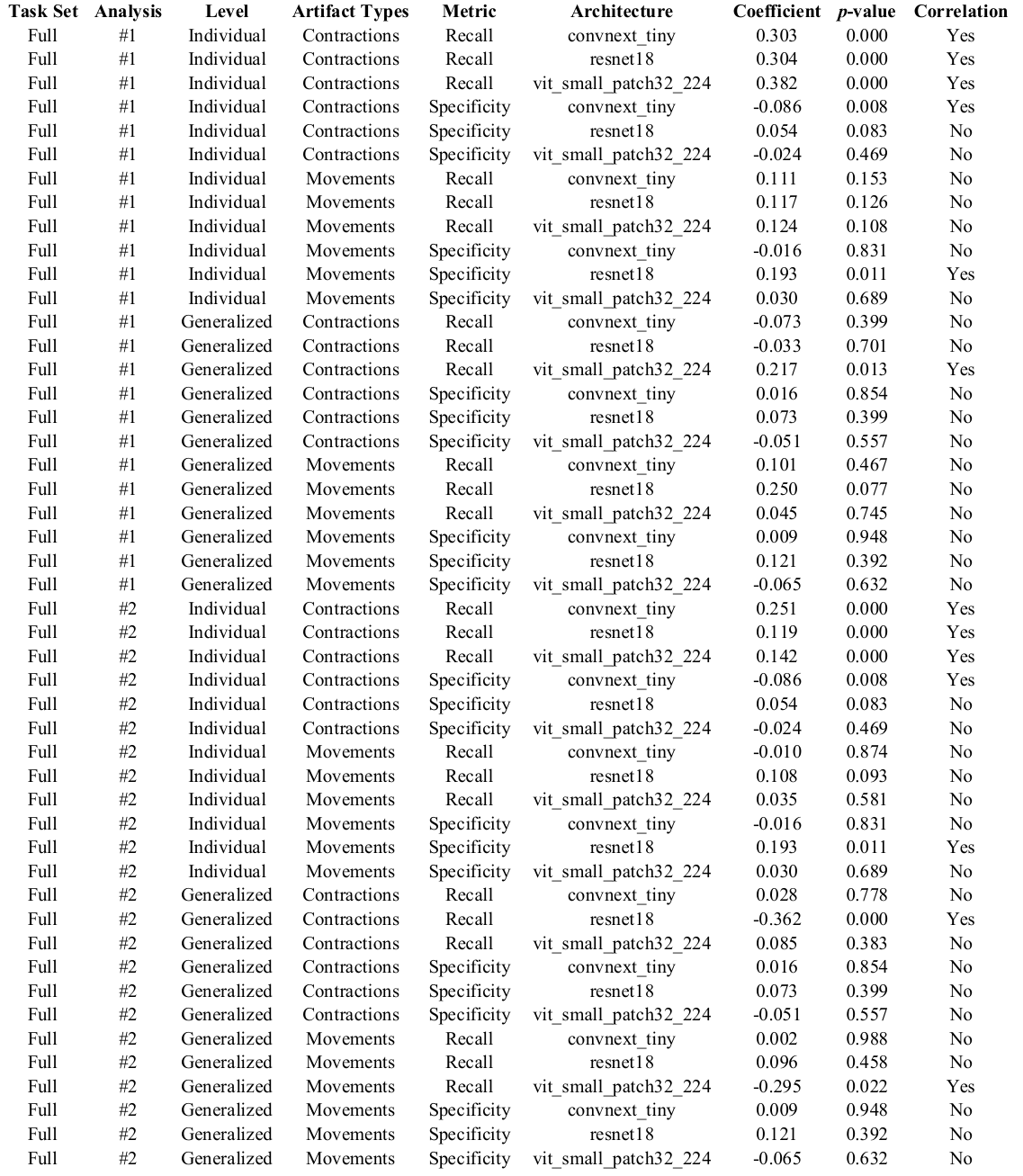} 
        \end{table*}

        \begin{table*} [!hbt]
            \centering
            \caption{Correlation coefficient and $p$-value of statistical tests measuring the correlation between the number of repetitions and performance with the selected set of tasks.}
            \label{tab_statistic_test_rep_corr_selected}
            \includegraphics[width=\linewidth]{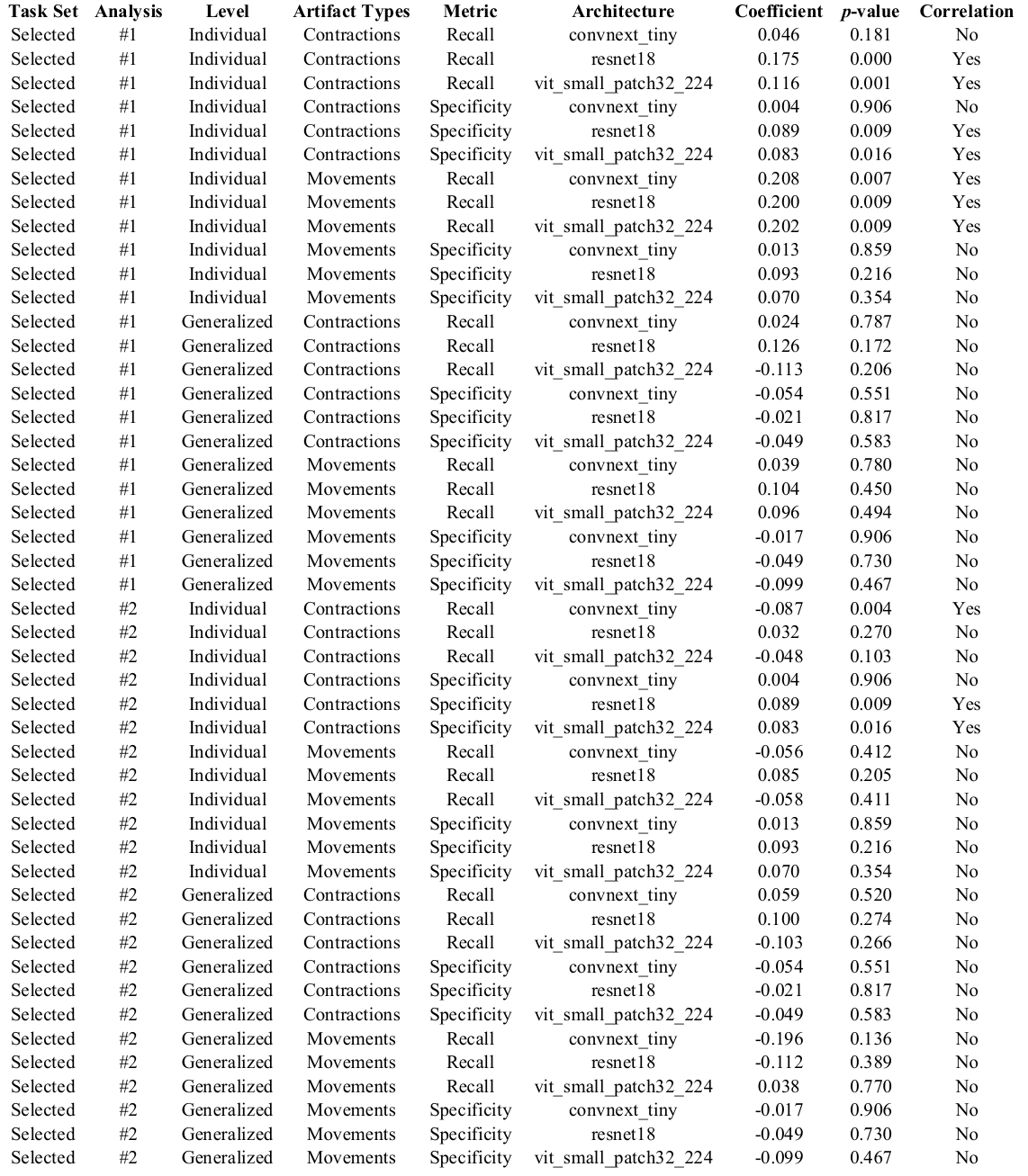} 
        \end{table*}
    
        \begin{table*}[!htb] 
            \centering 
            \caption{$p-$value of statistical tests comparing pre-training combined with calibration and individual models using all repetitions, all subjects and all three DL architectures.}
            \label{tab_statistic_test_pre_indi}
            \begin{tabularx}{\textwidth}{CCCCCC}
                 \bfseries Task Set & \bfseries Artifact Types  & \bfseries Analysis & \bfseries Level & \bfseries $p$-value of Recall  & \bfseries $p$-value of Specificity   \\ 
                 
                 Full               & Contractions              & \#3-(1) vs \#1 & Individual   & 0.375             & 0.060         \\ 
                 Full               & Contractions              & \#3-(2) vs \#2 & Individual   & 0.818             & 0.060         \\ 
                 Full               & Movements                 & \#3-(1) vs \#1 & Individual   & 0.042             & 0.244         \\ 
                 Full               & Movements                 & \#3-(2) vs \#2 & Individual   & 1.000             & 0.244         \\ 
                 Selected           & Contractions              & \#3-(1) vs \#1 & Individual   & 0.536             & 0.143         \\ 
                 Selected           & Contractions              & \#3-(2) vs \#2 & Individual   & 0.997             & 0.143         \\ 
                 Selected           & Movements                 & \#3-(1) vs \#1 & Individual   & 0.760             & 0.174         \\ 
                 Selected           & Movements                 & \#3-(2) vs \#2 & Individual   & 1.000             & 0.174         \\ 
            \end{tabularx}
        \end{table*}

        \begin{table*} [!hbt]
            \centering
            \caption{$p$-value of statistical tests measuring if the former architecture yields better results than the latter one in each architecture pair in one person, one model for analysis \#2.}
            \label{tab_statistic_test_best_result}
            \includegraphics[width=\linewidth]{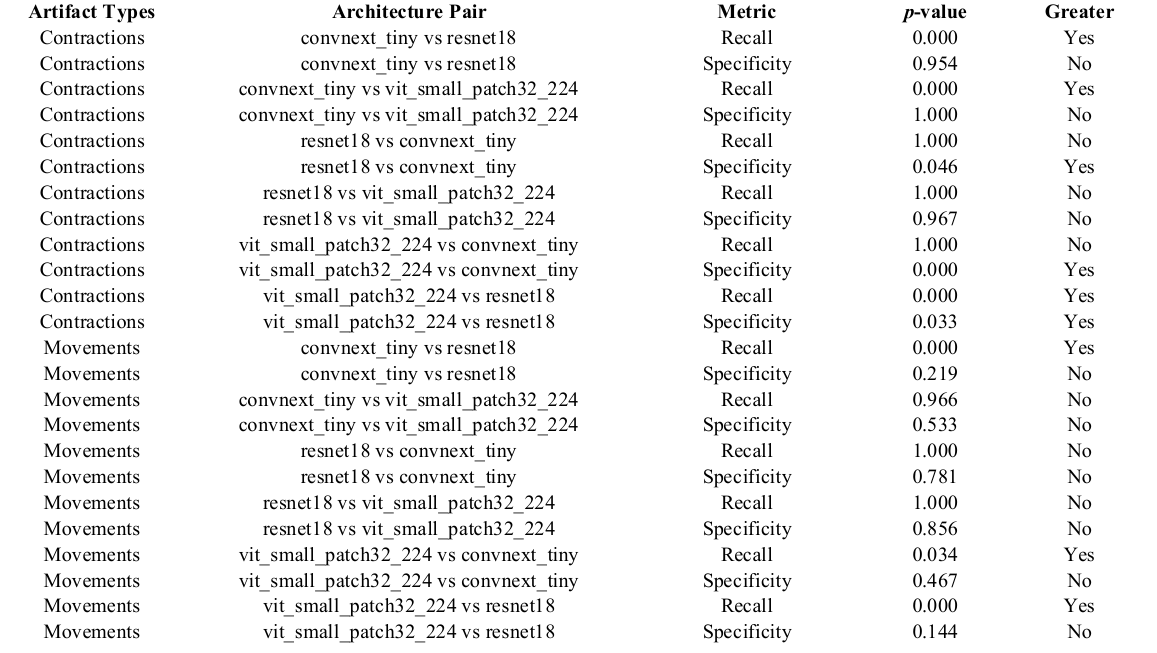} 
        \end{table*}
    
    \subsection{Performance and Misclassification With the Full Set of Tasks} \label{Appendix_before_kaout}
    
        The recall for isometric contractions for the one person, one model setting aligns with expectations, but the specificity for subjects 5, 6, 10, and 11 is notably lower (Tables \ref{tab_analysis1_recall_t_1p1m} and \ref{tab_analysis1_spec_t_1p1m}). This indicates substantial individual variability.
        
\begin{figure*}[!bt] 
            \centering
            \includegraphics[width= \linewidth]{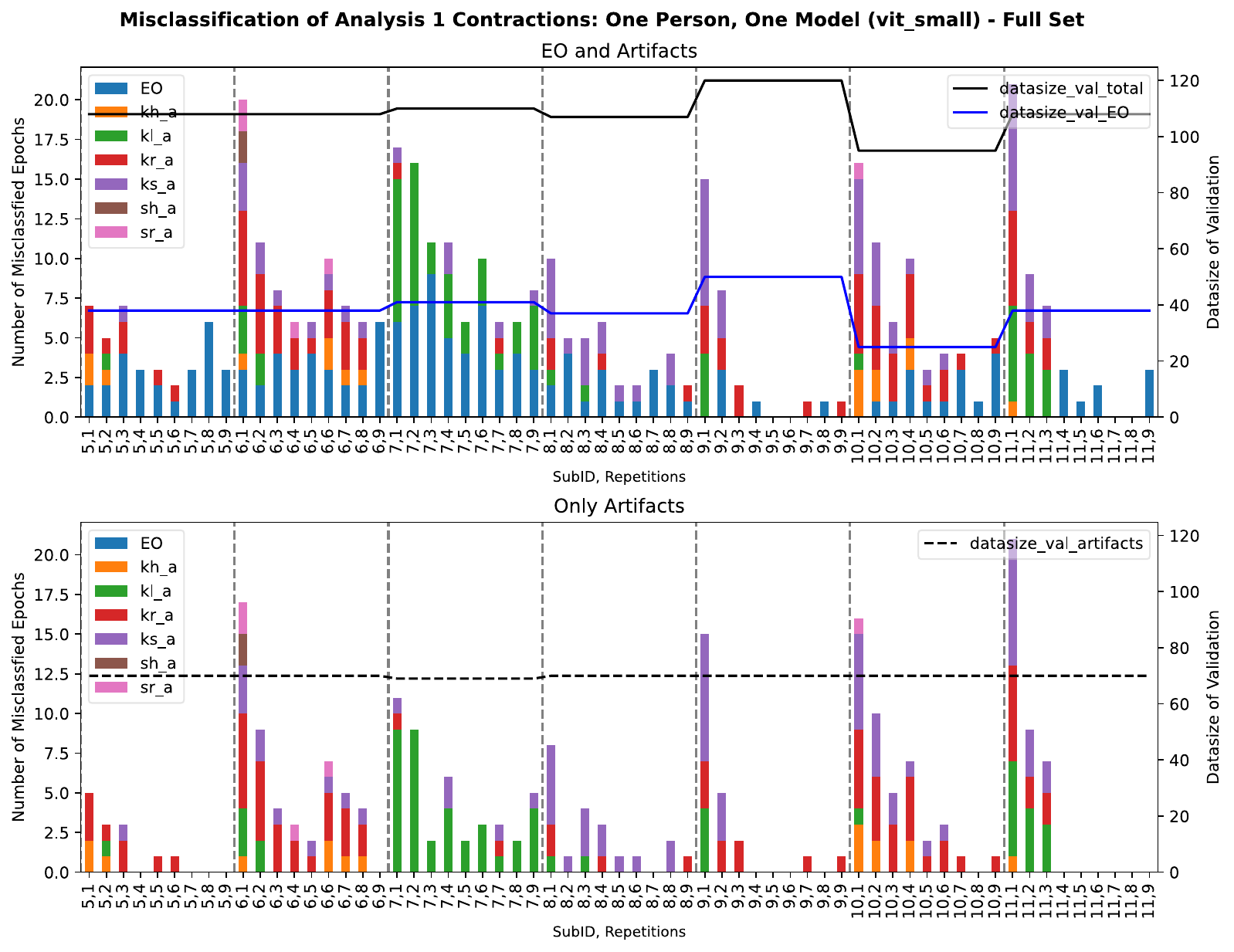}
            \caption{Misclassification of analysis \#1 isometric contractions: One person, one model using Vision Transformer Small. The x-axis shows the subject ID and the number of task repetitions. data\_size\_val\_total includes both EO and artifact epochs in the validation data. Artifacts are labeled with their IDs (refers to \autoref{tab_EMG_artifact_ID}).}
            \label{fig_misclass_expt1_indi_before}
        \end{figure*}

\begin{figure*}[!bt] 
            \centering
            \includegraphics[width= \linewidth]{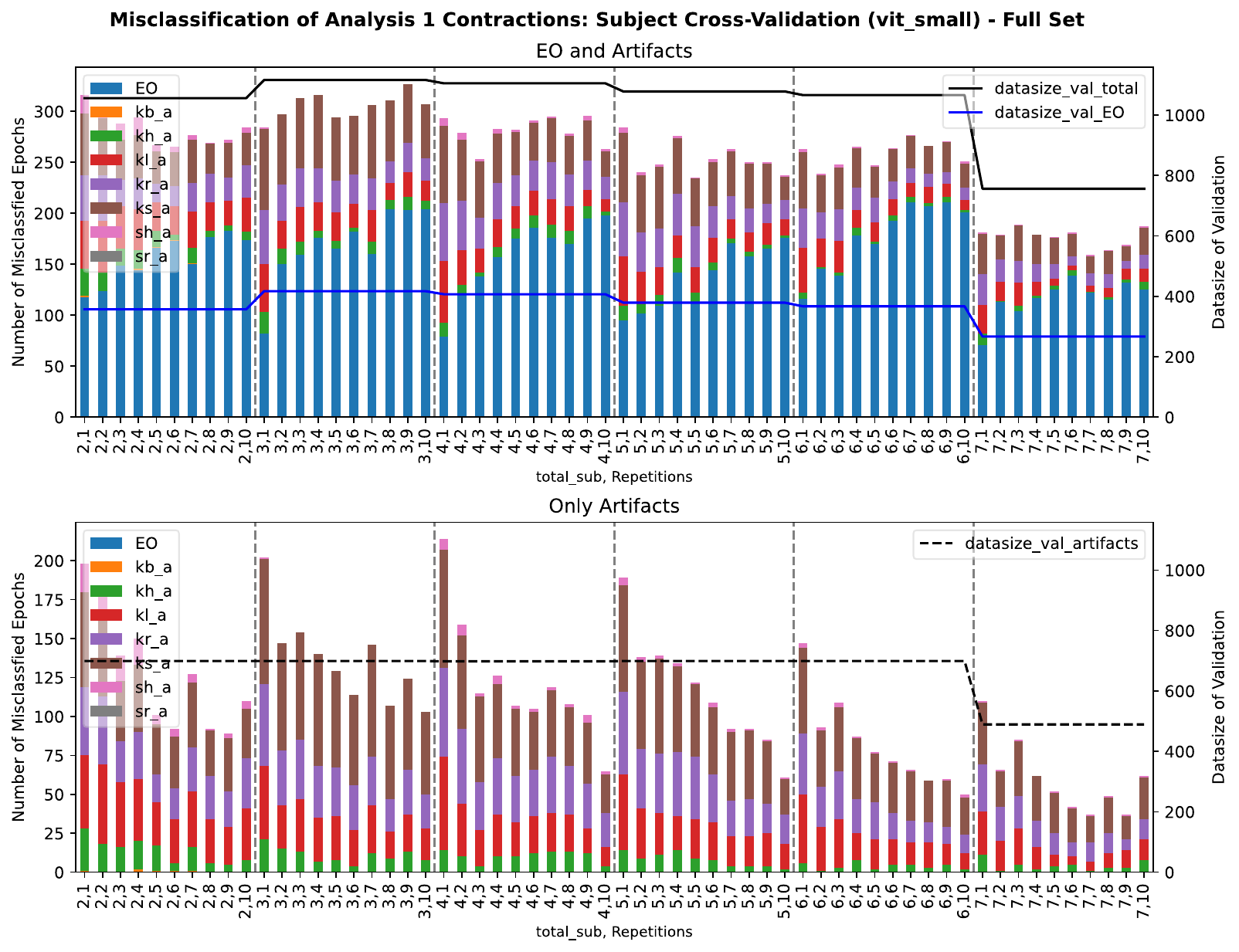}
            \caption{Misclassification of analysis \#1 isometric contractions: Subject cross-validation using Vision Transformer Small. The x-axis shows the number of total subjects and the number of task repetitions. data\_size\_val\_total includes both EO and artifact epochs in the validation data. Artifacts are labeled with their IDs (refers to \autoref{tab_EMG_artifact_ID}).}
            \label{fig_misclass_expt1_group_before}
        \end{figure*}

\begin{figure}[!hbt]
            \centering

            \subfloat[]{\includegraphics[width=\linewidth]{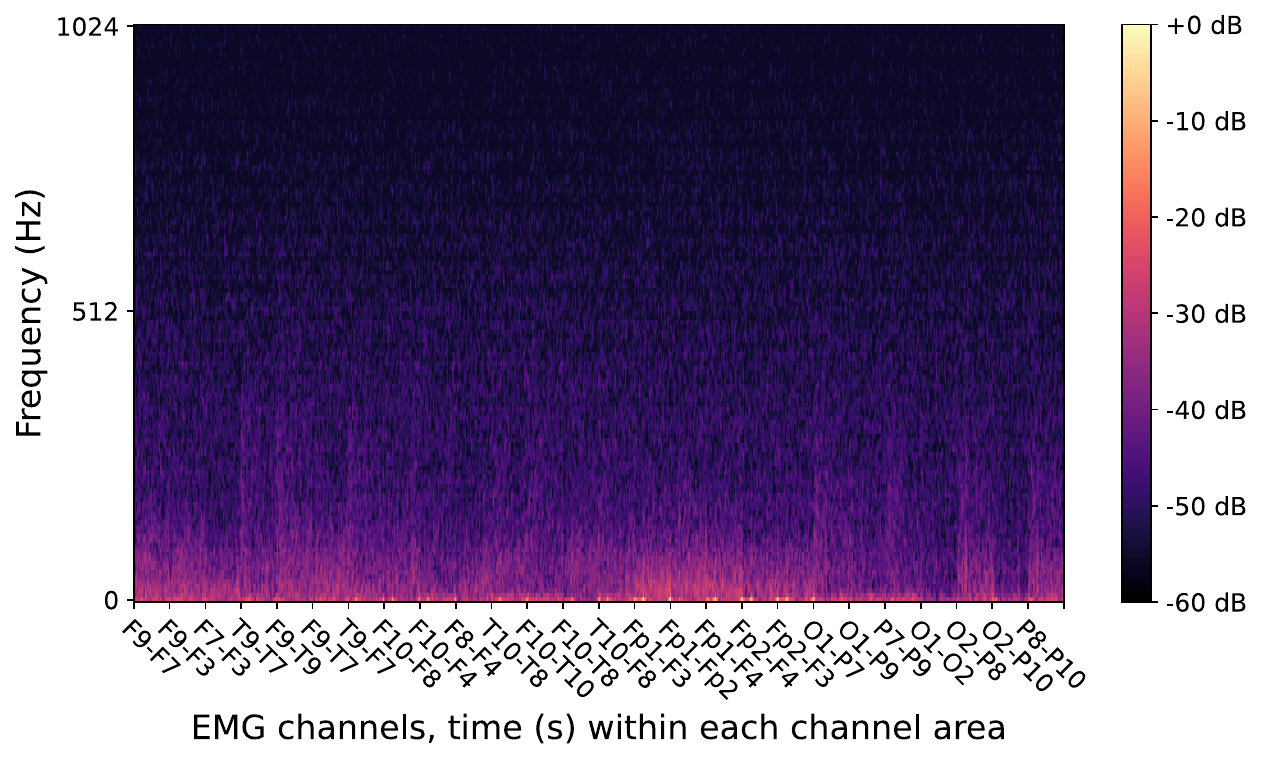}}\ 
            \subfloat[]{\includegraphics[width=\linewidth]{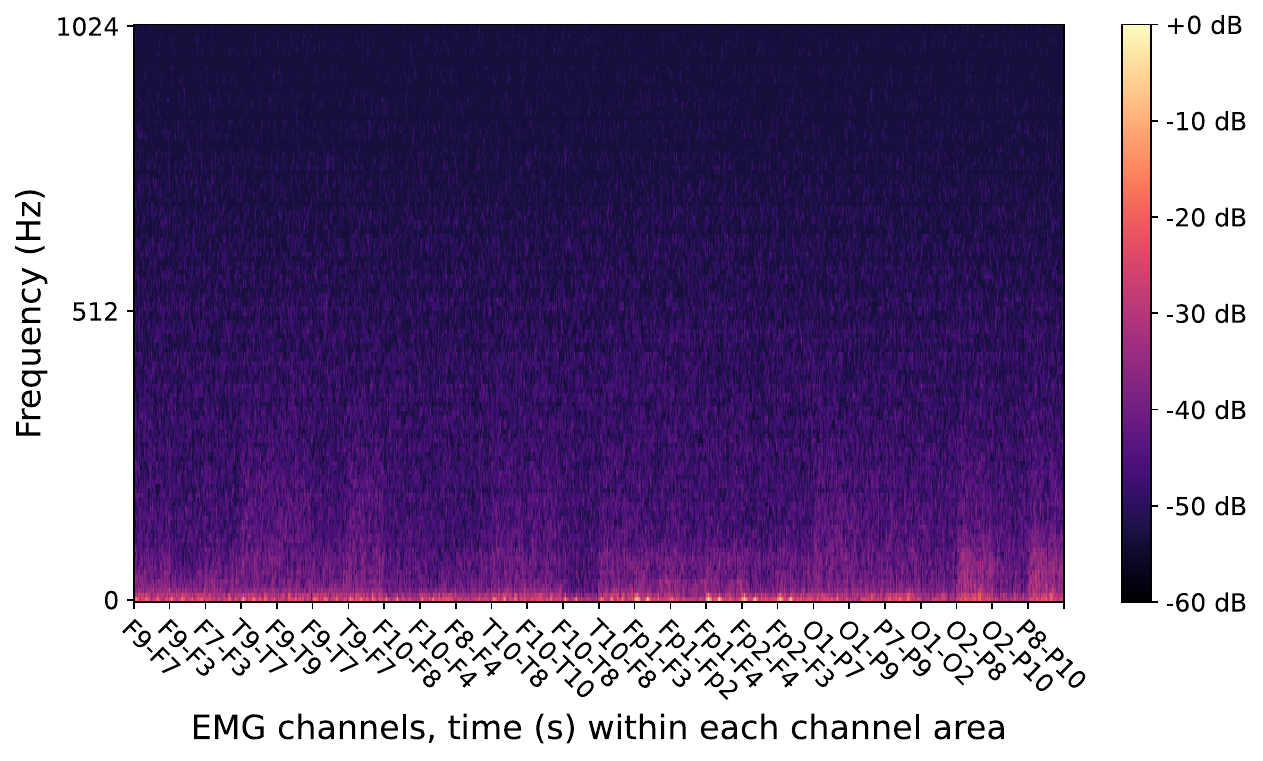}}
            \caption{(a) Illustration of a mel-spectrogram depicting an artifact epoch, particularly, the isometric contraction task involving turning the head left and holding. The EMG channels are concatenated along the x-axis, representing time within each channel area. (b) Exemplary mel-spectrogram showcasing a non-artifact epoch from EO recordings. (a) and (b) are from the same subject and closely resemble each other.}
            \label{fig_mel_spectro_misclass}
        \end{figure}

        The misclassified artifact epochs are predominantly the isometric contraction tasks related to the occipitalis muscle across all three architectures. The results for the Vision Transformer is given as an illustrative example (Figures \ref{fig_misclass_expt1_indi_before} and \ref{fig_misclass_expt1_group_before}). The mel-spectrograms of these artifact epochs often closely resemble those of non-artifact epochs (\autoref{fig_mel_spectro_misclass}). 
        
        In both one person, one model and subject cross-validation scenarios, an increase in repetition numbers results in a decrease in misclassified artifact epochs. However, in subject cross-validation, more non-artifact epochs are misclassified, especially among the four subjects mentioned in the one person, one model context. 
    
        \begin{table*}[!hbt]
            \centering
            \caption{Analysis \#1: Recall of isometric contractions in one person, one model with the full set of tasks. SubID means subject ID. Std means standard deviation.}
            \label{tab_analysis1_recall_t_1p1m}
            \includegraphics[width=\linewidth]{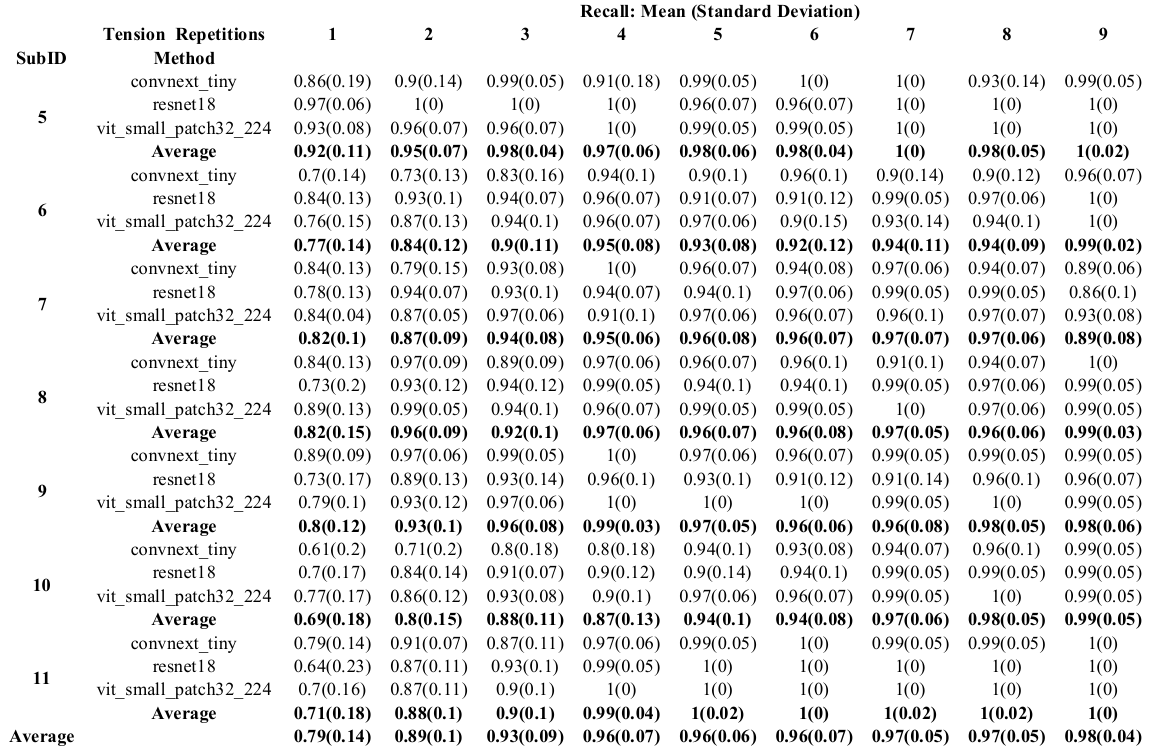} 
        \end{table*}
    
        \begin{table*}[!hbt]
            \centering
            \caption{Analysis \#1: Specificity of isometric contractions in one person, one model with the full set of tasks. SubID means subject ID. Std means standard deviation.}
            \label{tab_analysis1_spec_t_1p1m}
            \includegraphics[width=\linewidth]{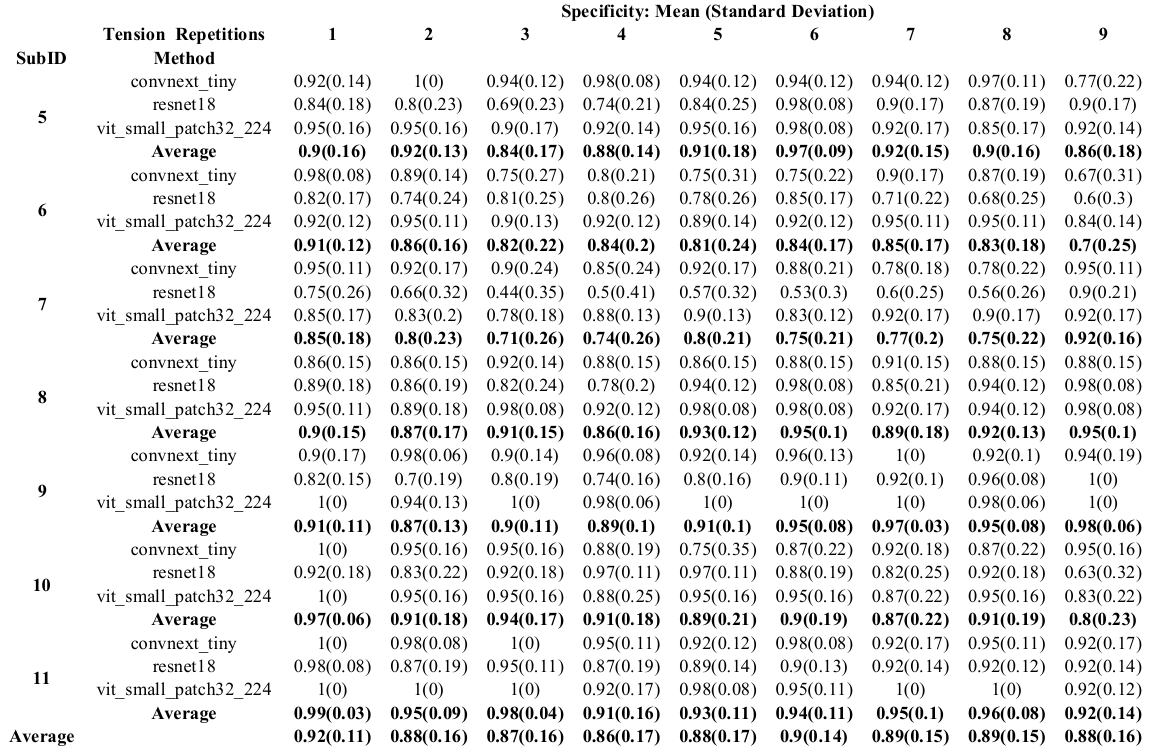} 
            
        \end{table*}

        \begin{table}[!hbt]
        \centering
        \caption{EMG artifact ID mapping}
        \label{tab_EMG_artifact_ID}
            \begin{tabular}{|c|c|c|} \hline 
                 \bfseries Artifact & \bfseries Artifacts & \bfseries Artifact Type\\
                 \bfseries IDs      &                     &                         \\ \hline \hline 
                 kb\_a& Jaw tensing & Isometric contractions \\ \hline 
                 kb\_db& Biting & Continuous movements\\ \hline 
                 kc\_db& Teeth grinding & Continuous movements \\ \hline 
                 sr\_a& Frowning & Isometric contractions \\ \hline 
                 sh\_a& Eyebrows raising and holding & Isometric contractions \\ \hline 
                 shr\_db& Eyebrows up and down & Continuous movements \\ \hline 
                 kl\_a& Head turning left and holding & Isometric contractions \\ \hline 
                 kr\_a& Head turning right and holding & Isometric contractions \\ \hline 
                 klr\_db& Head turning left and right & Continuous movements \\ \hline 
                 ks\_a& Head tilting downwards and holding & Isometric contractions \\ \hline 
                 kh\_a& Head tilting upwards and holding & Isometric contractions \\ \hline 
                 kn\_db& Nodding & Continuous movements \\\hline
            \end{tabular}
        \end{table}
        
    \subsection{Misclassification of Analysis \#2 with the Selected Set of Tasks}
        \label{Appendix_after_kaout}
\begin{figure*}
            \centering
            \includegraphics[width= \linewidth]{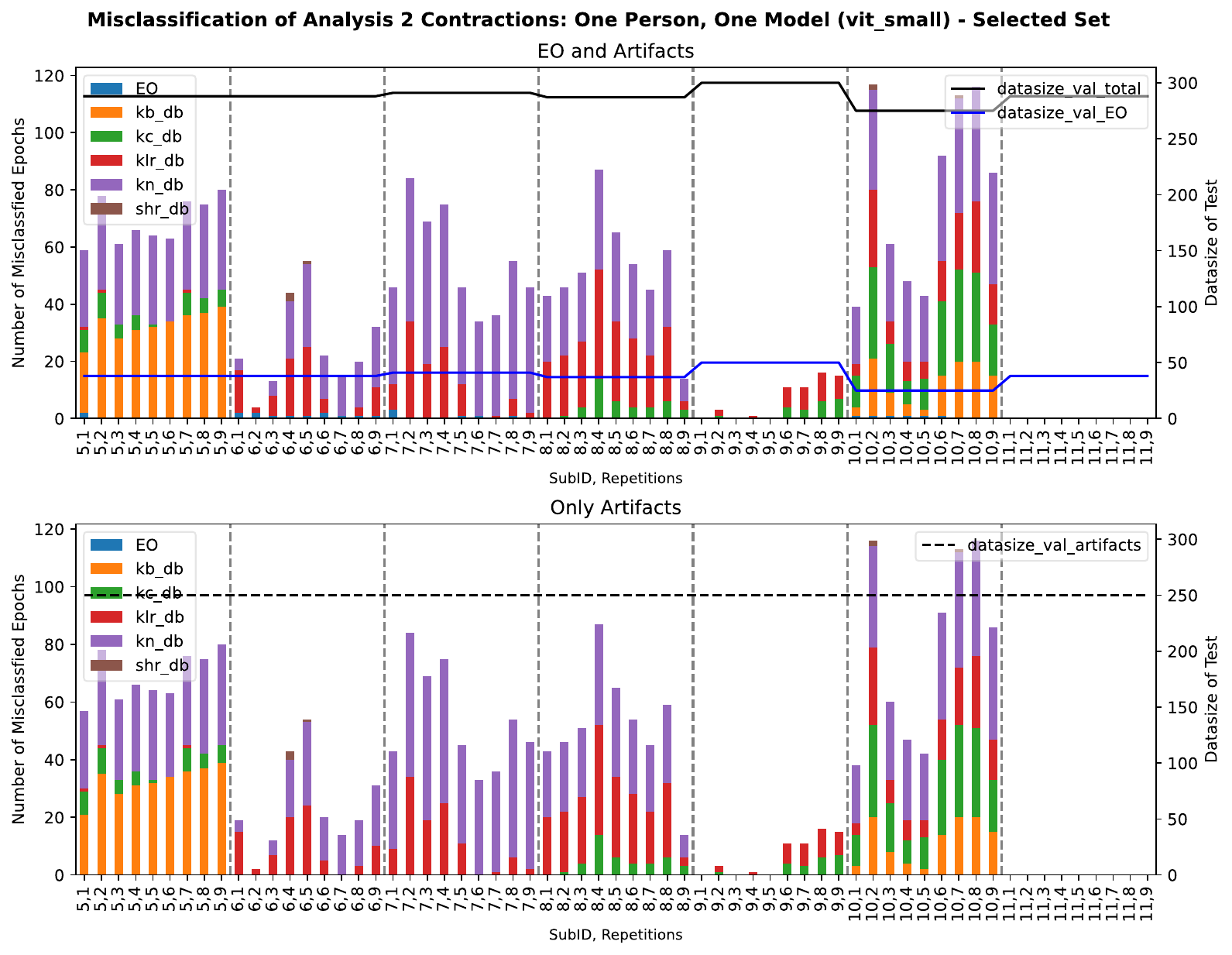}
            \caption{Misclassification of analysis \#2 isometric contractions: One person, one model using Vision Transformer Small. The x-axis shows the subject ID and the number of task repetitions. data\_size\_val\_total includes both EO and artifact epochs in the validation data. Artifacts are labeled with their IDs (refers to \autoref{tab_EMG_artifact_ID}). For subject eleven, both the recall and the specificity have reached one, therefore, no epoch is misclassified. }
            \label{fig_misclass_expt2_indi_after}
        \end{figure*}

        With the exclusion of the four isometric contraction tasks involving the occipitalis muscle, the performance of continuous movements slightly surpasses that of isometric contractions. This improvement is observed, because two continuous movement tasks targeting the occipitalis muscle (klr\_db and kn\_db) still remain in the validation data. These two movements are the most frequently misclassified artifacts (\autoref{fig_misclass_expt2_indi_after}).
        
\end{document}